\documentclass[a4paper,11pt]{article}
\pdfoutput=1
\usepackage{jcappub}
\usepackage{slashed}
\usepackage{mathtools}
\usepackage{graphicx}  
\usepackage{dcolumn}   
\usepackage{bm}        
\usepackage{amssymb,amsfonts,amsmath,physics}   
\usepackage{cancel}
\usepackage[normalem]{ulem}
\usepackage{tikz}

\newcommand{\Trh}{T_\text{rh}}
\newcommand{\Tmax}{T_\text{max}}
\newcommand{\arh}{a_\text{rh}}

\newcommand{\aend}{a_\text{end}}

\newcommand{\gs}{g_\star}
\newcommand{\gss}{g_{\star s}}

\newcommand{\rend}{\rho_\text{end}}
\newcommand{\rRH}{\rho_\text{rh}}
\newcommand{\Tqcd}{T_\text{qcd}}
\newcommand{\Tosc}{T_\text{osc}}
\newcommand{\rR}{\rho_R}

\newcommand{\Hrh}{H_\text{rh}}
\newcommand{\mueV}{\mu{\rm eV}}
\newcommand{\TBBN}{T_{\rm BBN}}
\newcommand{\DNeff}{\Delta N_{\rm eff}}
\newcommand{\fmax}{f_{\rm max}}
\newcommand{\ogw}{\Omega_{\rm GW}}
\newcommand{\matld}{\tilde{m}_a}

\title{Testing Axionic Dark Matter\\ during Gravitational Reheating}
\author[a]{Basabendu Barman,}
\author[b,c]{Arghyajit Datta}
\affiliation[a]{Department of Physics, School of Engineering and Sciences,
SRM University AP, Amaravati 522240, India}
\affiliation[b]{Center for Precision Neutrino Research, Chonnam National University, Gwangju 61186, Republic of Korea }
\affiliation[c]{Department of Physics, Kyungpook National University, Daegu 41566, Republic of Korea}
\emailAdd{basabendu.b@srmap.edu.in}
\emailAdd{arghyad053@gmail.com}
\abstract{Assuming axions are potential dark matter (DM) candidate that make up all of the DM abundance, we discuss production of axions via (i) standard misalignment mechanism during the period of reheating and (ii) graviton-mediated 2-to-2 scattering of the inflaton and bath particles, where the inflaton $\phi$ oscillates in a monomial potential $V(\phi)\propto\phi^k$ with a general equation of state. Considering reheating takes place purely gravitationally, mediated by massless gravitons, we explore the viable region of the parameter space that agrees with the observed DM relic abundance, satisfying bounds from big bang nucleosynthesis (BBN) and cosmic microwave background radiation (CMB). We also discuss complementarity between dedicated axion search experiments and futuristic gravitational wave search facilities in probing the viable parameter space.
}
\begin{document}
\maketitle
\section{Introduction}
\label{sec:intro}
The QCD axions~\cite{Weinberg:1977ma,Wilczek:1977pj}, pseudo-Nambu-Goldstone boson of the Peccei-Quinn (PQ) solution to the strong CP problem~\cite{Peccei:1977hh,Peccei:1977np,Peccei:1977ur} and the axion like particles (ALP), that could also arise from the spontaneous breaking of a global $U(1)$ symmetry, stand out as especially well motivated candidates for cold dark matter (DM)~\cite{Preskill:1982cy,Stecker:1982ws,Dine:1982ah,Abbott:1982af}. QCD axions and ALPs arise in various extensions of the Standard Model (SM) through spontaneous symmetry breaking (SSB) or from string compactification~\cite{Arias:2012az,Arvanitaki:2009fg}, with a  potential of getting discovered in the next decade (see, for example, Ref.~\cite{Adams:2022pbo}).

In the standard scenario, axions (by ``axions" we collectively refer to QCD axions and ALPs) can be produced in the early universe via the ``misalignment mechanism,"  in which the QCD axion or the ALP field modeled as the classical scalar field due to its bosonic nature and high occupation numbers, has a non-zero initial field value and non-zero potential energy, leading to oscillations of the field\footnote{Contrary to the conventional misalignment, the axion field may also possess a non-zero initial velocity in the so-called ``kinetic misalignment" mechanism~\cite{Co:2019jts,Chang:2019tvx,Barman:2021rdr}.}. For an $\mathcal{O}(1)$ initial misalignment angle $\theta_i$, the allowed mass window for QCD axion that produces the observed DM abundance turns out to be $10^{-6}\lesssim m_a\lesssim 10^{-5}$ eV, when the oscillation begins during radiation dominated Universe. For ALPs, on the other hand, the relic abundance depends on three parameters: the decay constant $f_a$, ALP mass $m_a$, and $\theta_i$, leading to strong bounds on the viable parameter space for $\theta_i\sim\mathcal{O}(1)$. It has been shown that deviations from standard cosmological histories can significantly broaden the parameter space for both axions and ALPs~\cite{Arias:2021rer,Bernal:2021bbv,Schiavone:2021imu,Arias:2022qjt}. Very recently, Ref.~\cite{Xu:2023lxw} has pointed out that such a conclusion also holds true if misalignment happens during the epoch of reheating if the inflaton\footnote{The QCD axion itself could have driven inflation as shown in~\cite{Freese:2004vs,Freese:2005kt,Pajer:2013fsa}.} $\phi$ oscillates in a monomial potential of the form $V(\phi)\propto\phi^k$, that provides a general equation of state (EoS) $0\leq w=(k-2)/(k+2)\lesssim 1$ for the inflaton.

On the other hand, the irrefutable coupling that one can imagine between matter particles (irrespective of dark and visible sector), is of gravitational origin, mediated by massless graviton~\cite{Choi:1994ax,Holstein:2006bh}. Such an interaction inevitably exists between the inflationary sector and axions. A graviton portal between the inflaton and the SM sector can also produce the radiation bath, completing the process of reheating even in the absence of direct coupling between the inflaton and the SM fields. This is dubbed as the ``gravitational reheating" scenario, that has been shown to be efficient for $k\gtrsim 9(4)$~\cite{Haque:2022kez,Barman:2022qgt}, considering (non-)minimal contribution to radiation. Now, $k>4$ implies a {\it stiff} EoS for the inflaton, that results in a significantly blue-tilted primordial gravitational wave (GW) spectrum, originating from the tensor perturbations during inflation~\cite{Giovannini:1998bp,Giovannini:1999bh,Riazuelo:2000fc,Seto:2003kc,Boyle:2007zx,Stewart:2007fu,Li:2021htg,Artymowski:2017pua,Caprini:2018mtu,Bettoni:2018pbl,Figueroa:2019paj,Opferkuch:2019zbd,Bernal:2020ywq,Ghoshal:2022ruy,Caldwell:2022qsj,Gouttenoire:2021jhk,Barman:2023ktz,Chakraborty:2023ocr}. Although a stiff period during reheating makes the GW signal potentially observable by GW detectors, but the very same enhancement also results in overproduction of GW energy density, that violates standard BBN and CMB predictions. 

Motivated by these, in this paper we have discussed a scenario where axion misalignment takes place in an inflaton dominated background during reheating, supplemented by an attractor potential for the inflaton. The reheating takes place purely gravitationally via scattering of the inflaton condensate into a pair of Higgs bosons, mediated by massless gravitons, in contrast to~\cite{Xu:2023lxw}, where reheating occurs via perturbative decay of the inflaton condensate into bosonic and fermionic final states. Interestingly, the oscillation temperature $\Tosc$ (and hence the relic abundance) becomes sensitive to the choice of $k$ (along with $m_a$ and $\theta_i$) in case of {\it minimal} gravitational reheating, and to both $\{k,\,\Trh\}$ in case where reheating takes place via a {\it non-minimal} coupling between the Higgs and gravity (Ricci scalar). We find that the standard misalignment mechanism during reheating opens up more parameter space for axions, compared to misalignment during radiation-domination, making up all of the DM abundance. The overproduction of the primordial GW energy density around the time of BBN rules out the minimal reheating scenario, while the non-minimal reheating remains within the sensitivity of future GW and axion search experiments. We thus find a complementarity between axion search and GW experimental facilities in constraining the allowed parameter space.

The paper is organized as follows. In Sec.~\ref{sec:model} we discuss the details of gravitational reheating and the generation of primordial gravitational wave. The mechanism of standard misalignment during reheating producing viable parameter space is discussed in Sec.~\ref{sec:qcd-axion}. Pure gravitational production of axions, mediated by graviton, is elaborated in Sec.~\ref{sec:grav-axion}. In Sec.~\ref{sec:expt} we discuss the discovery potential of this framework. Finally, we summarize our findings in Sec.~\ref{sec:concl}.

\section{Post-inflationary evolution of the Universe}
\label{sec:model}
The interaction between all matter fields and the gravitational field can be found by expanding the metric around Minkowski space-time $\eta_{\mu \nu}$ as $g_{\mu \nu}\simeq \eta_{\mu \nu}+\frac{2h_{\mu \nu}}{M_P}$, where $h_{\mu\nu}$ represents the canonically normalized quanta of the graviton. Consequently, one obtains possible gravitational interactions from the Lagrangian~\cite{Choi:1994ax,Holstein:2006bh}
\begin{align}
\sqrt{-g}\,\mathcal{L}_{\rm int}= -\frac{1}{M_P}\,h_{\mu \nu}\,\left(T^{\mu \nu}_{\rm SM}+T^{\mu \nu}_\phi + T^{\mu \nu}_X \right)\,,
\label{eq:grav-lgrng}
\end{align}
where ``SM" denotes the SM fields, $\phi$ is the inflaton while $X$ represents other beyond the SM (BSM) fields, which in our case is an axion. Here $M_P \simeq 2.45\times 10^{18}$ GeV is the reduced Planck mass. The form of stress-energy tensor $T^{\mu \nu}$ is dictated by the spin of the fields. For a generic spin-$0$ scalar $S$, such as the Higgs bosons, the inflaton or an axion\footnote{For axion this can be considered to be the the SM gauge singlet scalar field that breaks the PQ symmetry via SSB, and the angular direction of which corresponds to the axion field.}, it can be expressed as 
\begin{equation}\label{eq:tmunu}
T^{\mu \nu}_0 =
\partial^\mu S\,\partial^\nu S-
g^{\mu \nu}\,\left[\frac{1}{2}\partial^\alpha S\,\partial_\alpha S-V(S)\right]\,,
\end{equation}
where $V(S)$ represents the potential of the respective scalar.

For the inflationary potential, we adopt the following $\alpha$-attractor T-model~\cite{Kallosh:2013hoa} that provides a plateau region in the large field limit, leading to quasi-de Sitter expansion consistent with observation
\begin{equation}\label{eq:Vatt}
    V(\phi ) =\lambda\, M_P^4 \left[\tanh \left(\frac{\phi}{\sqrt{6\, \alpha}\, M_P}\right)\right]^k \simeq \lambda\, M_P^4 \times
    \begin{cases}
        1 & \; \text{for}\; \phi \gg M_P,\\
        \left(\frac{\phi}{\sqrt{6\,\alpha}\,M_P}\right)^k & \; \text{for}\; \phi\ll M_P\,.
    \end{cases}
\end{equation}
The overall scale of the potential, parameterized by the coupling $\lambda$, can be expressed in terms of the amplitude of the scalar perturbation power spectrum $A_S \simeq (2.1 \pm 0.1) \times 10^{-9}$~\cite{Planck:2018jri} as $\lambda \simeq \frac{18\,\pi^2\,\alpha\,A_S}{6^{k/2}\, N_\star^2} \,,$  where $N_\star$ is the number of $e$-folds measured from the end of inflation to the time when the pivot scale $k_\star = 0.05$~Mpc$^{-1}$ exits the horizon. Here onward we will also fix $\alpha=1/6$. The ending of inflation is marked when $\ddot a=0$, at which the inflaton field has a magnitude
\begin{align}
&\phi_e=\sqrt{\frac{3}{8}}M_P\ln\left[\frac{1}{2} + \frac{k}{3}\left(k + \sqrt{k^2+3}\right)\right]\,.
\label{eq:phiend}    
\end{align}
Furthermore, one can compute the effective mass of the inflaton field by taking the second derivative of the scalar potential, i.e., $m_\phi^2=\frac{\partial^2 V}{\partial\phi^2}$ that at the end of inflation turns out to be $m_\phi(a=\aend)\simeq 10^{13}$ GeV for above choice of the CMB observable, with a very mild dependence on the exponent $k$. With this set-up we will now move on to the computation of gravitational reheating temperature.

\subsection{Gravitational reheating}
\label{sec:grav-reheat}
In order to track the evolution of the inflaton $(\rho_\phi)$ and the radiation $(\rR)$ energy densities during reheating, we solve the following set of coupled Boltzmann equations (BEQ)~\cite{Giudice:2000ex}
\begin{align}
    &\frac{d\rho_\phi}{dt} + 3H\,(1+w_\phi)\, \rho_\phi = -(1+w_\phi)\,\Gamma_\phi\, \rho_\phi\,,\label{eq:BErhop}\\
    &\frac{d\rho_R}{dt} + 4H\, \rho_R = + (1+w_\phi)\,\Gamma_\phi\, \rho_\phi\,, 
    \label{eq:BErhoR}
\end{align}
together with 
\begin{align}
H = \sqrt{\frac{\rho_\phi+\rR}{3\,M_P^2}}\,.    
\end{align}
Here, $w_\phi\equiv\frac{p_\phi}{\rho_\phi}=\frac{k-2}{k+2}$~\cite{Garcia:2020wiy} is the general equation of state parameter for the inflaton. Since during most part of the reheating, the total energy density is dominated by the inflaton, the expansion rate corresponding to the term $3\,H\,(1+w_\phi)\,\rho_\phi$ dominates over the reaction rate $(1+w_\phi)\,\Gamma_\phi\rho_\phi$ in Eq.~\eqref{eq:BErhop}. As a consequence, one can solve Eq.~\eqref{eq:BErhop} analytically by neglecting the right-hand side and obtain 
\begin{equation}
\rho_\phi(a) \simeq \rho_{\rm end}\,\left(\frac{a_{\rm end}}{a} \right)^\frac{6\,k}{k+2}\,,
\label{eq:rhoPh}
\end{equation}
with corresponding Hubble rate
\begin{align}\label{eq:hubble}
& H(a)\simeq H_{\rm end}\,\left(\frac{a_{\rm end}}{a} \right)^\frac{3\,k}{k+2}\,.    
\end{align}  
Solution of Eq.~\eqref{eq:BErhoR} requires the information of the reaction rates of inflaton. Since we are interested in the purely gravitational reheating, i.e., no direct coupling between the inflaton and the SM states is present, the production rate of radiation in that case can be evaluated as~\cite{Clery:2021bwz, Clery:2022wib, Co:2022bgh}
\begin{equation}
   (1+w_\phi)\,\Gamma_\phi\, \rho_\phi = R^{\phi^k}_H \simeq \frac{N_h\rho_{\phi}^2}{16\pi M_P^4} \sum_{n=1}^{\infty}  2n\omega|{\mathcal{P}}^k_{2n}|^2 = \alpha_k M_P^5 \left(\frac{\rho_{\phi}}{M_P^4}\right)^{\frac{5k-2}{2k}} \, ,
   \label{Eq:ratephik}
\end{equation}
by considering the graviton propagator for momentum $q$ as
\begin{equation}
  \Pi^{\mu\nu\rho\sigma}(p) = \frac{\eta^{\rho\nu}\eta^{\sigma\mu} + 
 \eta^{\rho\mu}\eta^{\sigma\nu} - \eta^{\rho\sigma}\eta^{\mu\nu} }{2\,q^2}\,.
\end{equation}
Here, we have considered the interaction of inflaton with only the SM Higgs field, neglecting its mass. consequently, $N_h=4$ is considered as the number of internal degrees of freedom for one complex Higgs doublet. While evaluating the interaction rate, the inflaton is treated as a time-dependent external and classical background field, which we parametrize as
\begin{equation}
    \label{Eq:oscillation}
    \phi(t)= \phi_0(t)\times\mathcal{T}(t) = \phi_0(t)\sum_{n=-\infty}^{\infty}\,{\cal T}_n\,e^{-in \omega t}\,,
\end{equation}
where $\phi_0(t)$ is the time-dependent amplitude that includes the effects of redshift and $\mathcal{T}(t)$ describes the periodicity of the oscillation. We also expand the potential energy in terms of the Fourier modes as~\cite{Ichikawa:2008ne,Kainulainen:2016vzv,Clery:2021bwz,Co:2022bgh,Garcia:2020wiy,Ahmed:2022qeh}
\begin{align}
V(\phi)=V(\phi_0)\sum_{n=-\infty}^{\infty} {\cal P}_n^ke^{-in \omega t}
=\rho_\phi\sum_{n = -\infty}^{\infty} {\cal P}_n^ke^{-in \omega t}\,,
\end{align}
where the frequency of oscillations of $\phi$ field reads~\cite{Garcia:2020wiy}
\begin{equation}
\label{eq:angfrequency}
\omega=m_\phi \sqrt{\frac{\pi k}{2(k-1)}}
\frac{\Gamma(\frac{1}{2}+\frac{1}{k})}{\Gamma(\frac{1}{k})}\,.
\end{equation}
By solving Eq.~\eqref{eq:BErhoR}, using the Eq.~\eqref{eq:rhoPh} together with the interaction rate~\eqref{Eq:ratephik}, one obtains the radiation energy density as
\begin{align}
    \rho_R(a) \simeq\sqrt{3}\,\alpha_k\,M_P^4\, \left(\frac{k+2}{8k-14}\right) \left(\frac{\rho_{\rm end}}{M_P^4}\right)^{\frac{2k-1}{k}} \left(\frac{a_{\rm end}}{a}\right)^4\left[1-\left(\frac{a_{\rm end}}{a}\right)^{\frac{8k-14}{k+2}}\right].
    \label{eq:rhoR}
\end{align}
One can relate this energy density with the bath temperature via
\begin{align}
    \rho_R(a)= \frac{\pi^2 g_*}{30}\, T^4(a),
    \label{eq:t}
\end{align}
where $g_*$ is the relativistic degrees of freedoms present in the thermal bath, and we assume instantaneous thermalization. From Eq.~\eqref{eq:BErhop} and \eqref{eq:BErhoR}, we note that the thermalization process of the SM particles produced from the inflaton scattering helps the Universe to attain a  maximum temperature $\Tmax$ right at the end of inflation. Subsequently the temperature falls to $\Trh$, where equality between $\rho_\phi$ and $\rho_R$ is achieved. As a result, reheating temperature can be evaluated as
\begin{align}\label{eq:grav-trh}
& \Trh^4 = \frac{30}{\pi^2\,g_{\rm RH}}\, M_P^4\,\left(\frac{\rend}{M_P^4}\right)^{\frac{4k-7}{k-4}}\,\left(\frac{\alpha_k\,\sqrt{3}\,(k+2)}{8k-14}\right)^\frac{3k}{k-4}\,. 
\end{align}
One can further note that, for $\aend\ll a\ll\arh$,  the temperature evolves as [cf. Eq.~\eqref{eq:rhoR}]
\begin{align}
& T(a)=\Trh\,\left(\frac{\arh}{a}\right)\,.
\end{align}
Purely gravitational scattering process discussed above may not always lead to efficient reheating of our Universe. For example, with $k = 2$, the radiation energy density produced by inflaton scattering never comes to dominate the energy density and can not lead to reheating. For $k > 4$ reheating from gravitational scattering is possible. However it is very inefficient,i.e., with $k=6$, from Eq.~\eqref{eq:grav-trh}, we obtain $\Trh\ll 1~\text{eV}< T_{\rm BBN}$. One actually needs to go beyond $k=9$, for which the gravitational reheating temperature can be found to be $\Trh \simeq 2$ MeV. On top of that, as we shall explain, the minimal gravitational reheating scenario is completely ruled out from the overproduction of inflationary GW energy density around the time of BBN. This motivates us to go beyond the minimal scenario and introduce a non-minimal coupling. 

The non-minimal coupling of Higgs bosons to gravity (via an interaction of the form $\xi_h\,H^\dagger H\,\mathcal{R}$, $\mathcal{R}$ being the Ricci scalar~\cite{Clery:2022wib}) provides an additional channel to reheat with the rate~\cite{Clery:2022wib,Co:2022bgh}
\begin{equation}
   (1+\omega_\phi)\,\Gamma_\phi= R^{\phi,\xi_h}_H \simeq \frac{\xi_h^2N_h}{4\pi M_P^4} \sum_{n=1}^{\infty}  2n\omega \left|2\times{\mathcal{P}}^k_{2n}\rho_{\phi} + \frac{(n\omega)^2}{2}\phi_0^2\,|\mathcal{T}_n|^2 \right|^2 = \alpha_k^{\xi_h}\,M_P^5\,\left(\frac{\rho_{\phi}}{M_P^4}\right)^{\frac{5k-2}{2k}}\,,
   \label{ratexi}
\end{equation}
where numerical estimates of the co-efficient $\alpha_k^{(\xi_h)}$ for different values of $k$ are reported in Table. 1 of Ref.~\cite{Barman:2022qgt}. The non-minimal reheating temperature in this case can be obtained as 
\begin{align}\label{eq:non-minimal-trh}
& \left(\Trh^{\xi_h}\right)^4 = \frac{30}{\pi^2\,g_{\rm RH}}\, M_P^4\,\left(\frac{\rend}{M_P^4}\right)^{\frac{4k-7}{k-4}}\,\left(\frac{\alpha_k^{\xi_h}\,\sqrt{3}\,(k+2)}{8k-14}\right)^\frac{3k}{k-4}\,,
\end{align}
while the maximum temperature in this case is determined by 
\begin{align}\label{eq:tmax-xi}
& \rho_\text{max}^\xi\simeq \sqrt{3}\,\alpha_k^{\xi_h}\,M_P^4\, \left(\frac{\rend}{M_P^4}\right)^\frac{2k-1}{k}\,\frac{k+2}{12k-16}\,\left(\frac{2k+4}{6k-3}\right)^\frac{2k+4}{4k-7} \equiv \frac{\pi^2}{30}\,g_\star\,\left(\Tmax^{\xi_h}\right)^4 \,,
\end{align}
which is $\mathcal{O} (10^{12})$ GeV, with mild dependence on $k$~\cite{Clery:2021bwz,Barman:2022qgt}. With non-minimal contribution taken into account, we note, reheating can be completed before the onset of BBN for $k>4$, by tuning the non-minimal coupling properly. 
\begin{figure}[t!]
    \centering
    \includegraphics[scale=0.36]{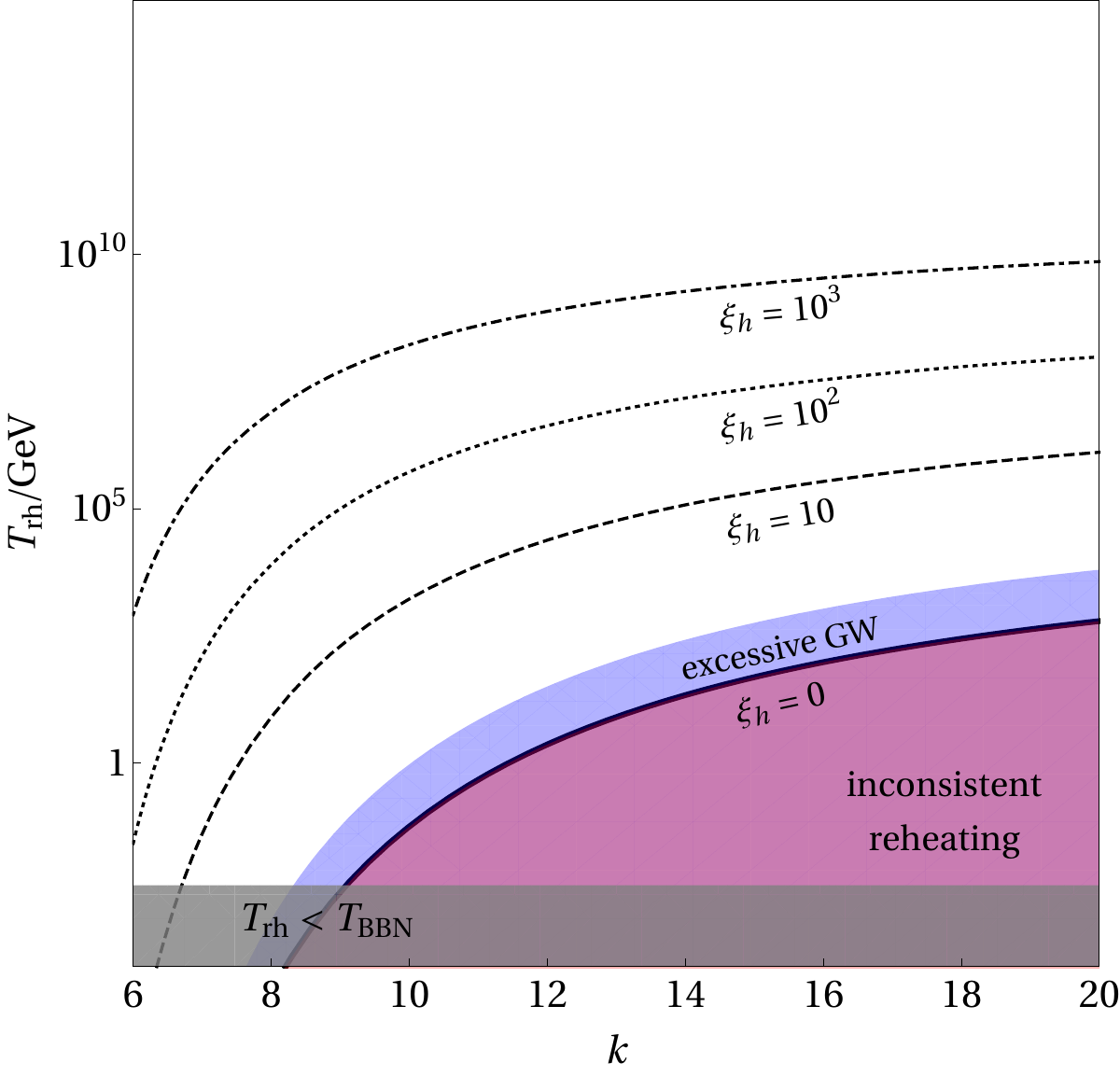}~~~~\includegraphics[scale=0.35]{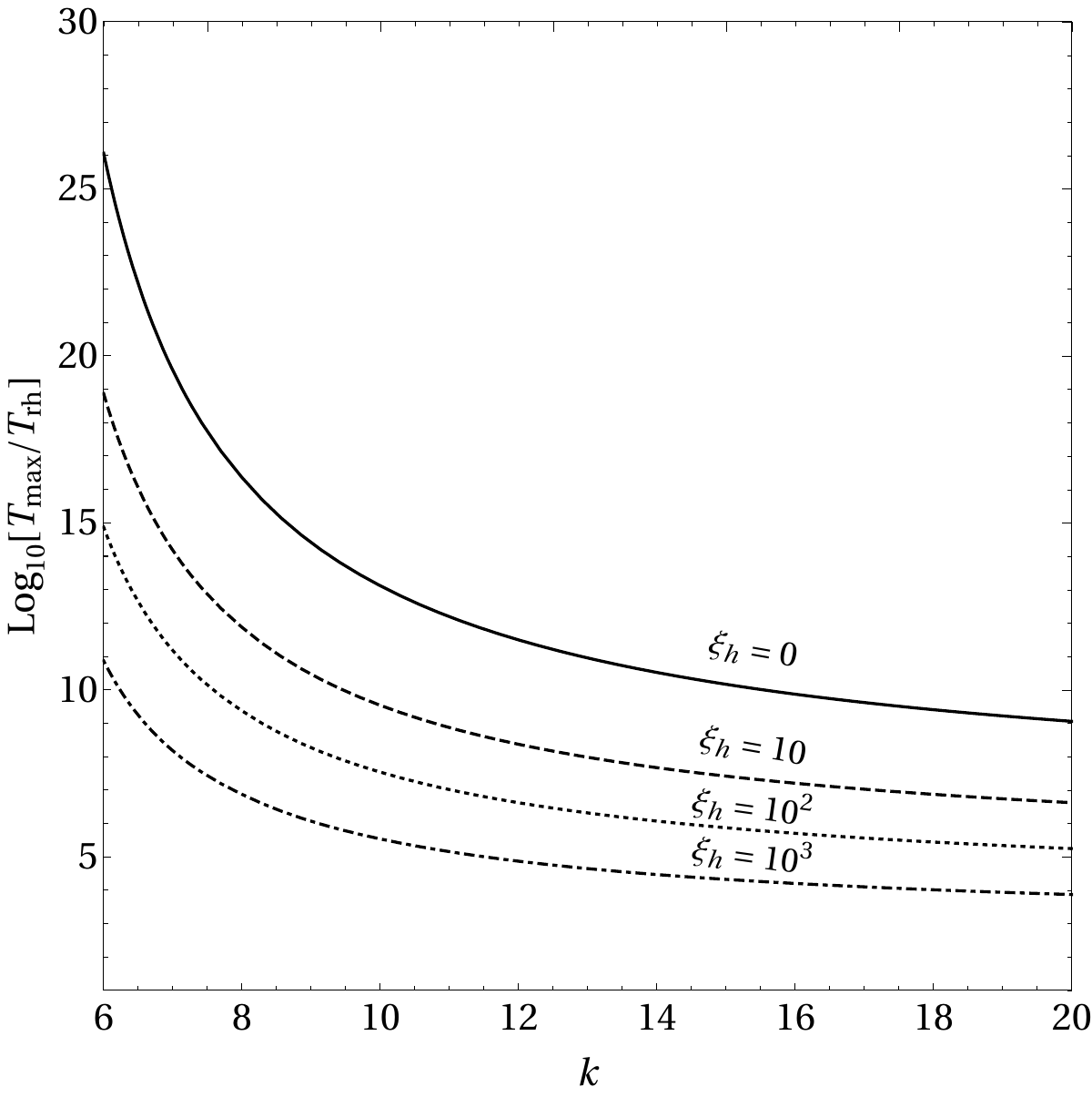}
    \caption{{\it Left:} Gravitational reheating temperature as function of $k$, where each curve corresponds to a fixed non-minimal coupling $\xi_h$, as indicated in the plot. The shaded regions are forbidden from $\DNeff$ bound due to Planck from excessive production of primordial GW as discussed in subsection.~\ref{sec:pgw}, ruling out the minimal gravitational reheating scenario $\xi_h=0$ (solid black curve). {\it Right:} Ratio of $\Trh$ to $\Tmax$, as a function of $k$ for the same choice of non-minimal couplings as in the left panel.}
    \label{fig:Trh}
\end{figure} 

We show reheating temperature as a function of $k$ for different choices of the non-minimal coupling in the left panel of Fig.~\ref{fig:Trh}, where the black solid curve corresponds to the minimal gravitational reheating scenario $(\xi_h=0)$. The shaded region, corresponding to minimal gravitational reheating, is ruled out from GW overproduction as we shall elaborate in subsection.~\ref{sec:pgw}. The right panel shows corresponding maximum temperature $\Tmax$ for each $\xi_h$. For higher $\xi_h$, as expected, the ratio $\Tmax/\Trh$ becomes smaller since $\Trh$ can be larger in those cases. Thus, for a given $k$, thanks to the non-minimal coupling, it is possible to have $\Trh$ much larger than that compared to minimal gravitational reheating scenario. Hereafter, for the non-minimal case, we will consider $\{k,\,\Trh\}$ as free-parameters.  

\subsection{Inflationary gravitational wave}
\label{sec:pgw}
In this section we briefly describe the background theory for computing the spectral energy density of primordial gravitational wave (GW), emerging from the tensor fluctuations during inflation (for a review, see, for example, Ref.~\cite{Caprini:2018mtu}). GWs are transverse ($\partial_ih_{ij} = 0$) and traceless ($h_{ii} = 0$) metric perturbations $ds^2 = a^2(t)\,(-dt^2+(\delta_{ij}+h_{ij})\,dx^i\,dx^j)$. Their energy density spectrum per momentum mode (at sub-horizon scales) is defined as~\cite{Boyle:2005se,Caprini:2018mtu}
\begin{align}
\Omega_{\text{GW}}(t, k) \equiv\dfrac{1}{\rho_{\text{crit}}}\dfrac{d\rho_{\text{GW}}(t,k)}{d\ln k} =\frac{k^2}{12\,a^2(t)\,H^2(t)}\,\Delta_h^2(t,k)\,,  \label{eq:GWenergySpectrum}
\end{align}
where $\Delta_h^2(t,k)$ is the tensor power spectrum at arbitrary times, defined as
\begin{align}
\left<h_{ij}(t,\mathbf{x})h^{ij}(t,\mathbf{x})\right>&\equiv\int \frac{dk}{k} \Delta_h^2(t,k)\,,
\end{align}
with $\left<...\right>$ denoting an average over a statistical ensemble. One can factorize the tensor power spectrum as~\cite{Watanabe:2006qe,Saikawa:2018rcs}
\begin{align}
\Delta_h^2(t,k)\equiv T_h(t,k) \Delta_{h,\text{inf}}^2(k)\,, \label{transferfunc}
\end{align}
with 
\begin{align}
T_h(t,k)=\frac{1}{2}\,\left(\frac{a_{\rm hc}}{a}\right)^2\,,  
\end{align}
the transfer function~\cite{Boyle:2005se,Saikawa:2018rcs,Caprini:2018mtu} (1/2 appears due to oscillation-averaging the tensor mode functions) and $ \Delta_{h,\text{inf}}^2(k)$ representing the primordial tensor spectrum from inflation~\cite{Caprini:2018mtu}
\begin{align}
\Delta_{h,\text{inf}}^2(k) \simeq 
\frac{2}{\pi^2}\,\left(\frac{H_{\rm inf}}{M_P}\right)^2\,\left(\frac{k}{k_p}\right)^{n_t}\,,
\label{eq:InfSpectrum}
\end{align}
where $n_t$ is the spectral tilt, $k_{p}$ denotes a pivot scale of the order Hubble rate at the time of CMB decoupling and $H_{\rm inf}$ represents the Hubble rate when the mode $k_p$ exited the Hubble radius during inflation. For simplicity, we shall assume an exact scale-invariant inflationary spectrum or, in other words, $n_t=0$. The GW background modes produced from such tensor fluctuations can cross outside the horizon during inflation when $k < a\,H$ holds and can be considered as classical modes. Subsequently, at a later time after inflation, these modes reenter the horizon $(k > a\,H)$ and form a stochastic background. 

Now, any extra radiation component, in addition to those of the SM, can be quantified in terms of the $\DNeff$. Since GW energy density scales the same way as that of free radiation, it is possible to put an upper bound on $\rho_\text{GW}$ as an extra radiation component at the time of BBN and/or CMB decoupling. This can be done by computing the total radiation energy density in the late Universe as
\begin{equation}
    \rho_\text{rad}(T\ll m_e) = \rho_\gamma + \rho_\nu + \rho_\text{GW} = \left[1 + \frac78 \left(\frac{T_\nu}{T_\gamma}\right)^4 N_\text{eff}\right] \rho_\gamma\,,
\end{equation}
where $\rho_\gamma$, $\rho_\nu$, and $\rho_\text{GW}$ correspond to the photon, SM neutrino, and GW energy densities, respectively, with $T_\nu/T_\gamma = (4/11)^{1/3}$. Within the SM, taking the non-instantaneous neutrino decoupling into account, one finds $N_\text{eff}^\text{SM} = 3.044$~\cite{Dodelson:1992km, Hannestad:1995rs, Dolgov:1997mb, Mangano:2005cc, deSalas:2016ztq, EscuderoAbenza:2020cmq, Akita:2020szl, Froustey:2020mcq, Bennett:2020zkv}, while the presence of GW results in a modification 
\begin{align}
& \DNeff = N_\text{eff}-N_\text{eff}^\text{SM} = \frac{8}{7}\,\left(\frac{11}{4}\right)^\frac{4}{3}\,\left(\frac{\rho_\text{GW}(T)}{\rho_\gamma(T)}\right)\,.
\end{align}
The above relation can be utilized to put a constraint on the GW energy density red-shifted to today via~\cite{Maggiore:1999vm,Boyle:2007zx,Caprini:2018mtu}
\begin{align}\label{eq:GW-neff}
& \left(\frac{h^2 \rho_\text{GW}}{\rho_c}\right)\Bigg|_0=
\int_{f_\text{BBN}}^{\fmax}\,\frac{df}{f}\,h^2\,\ogw^{(0)}(f)
\leq\frac{7}{8}\,\left(\frac{4}{11}\right)^\frac{4}{3}\,\Omega_\gamma^{(0)}\,h^2 \DNeff ,
\end{align}
which leads to $\ogw^{(0)}\,h^2\simeq 5.62\times 10^{-6}\,\DNeff$ where, $\Omega_\gamma^{(0)}\,h^2\simeq 2.47\times 10^{-5}$ is the relic photon abundance at the present epoch while $f=k/(2\pi a_0)$ is the present day frequency of the physical wave number $k$. Here $f_\text{max}$ corresponds to maximum frequency that re-enter the horizon right after the end of inflation when $k_{\rm max}=\aend\,H_{\rm end} $, and is given by
\begin{align}
&  \fmax = \frac{H(\Tmax)}{2\pi}\,\frac{\aend}{a_0}\,,
\end{align}
while $f_{\rm BBN}\simeq 2\times 10^{-11}$ Hz, corresponds to the mode $k_{\rm BBN}=a_{\rm BBN}\,H_{\rm BBN}$ at the
time of BBN. Using the present CMB measurement from Planck legacy data~\cite{Planck:2018jri}, we find $\ogw^{(0)}\,h^2\lesssim 2\times 10^{-6}$, with $\DNeff\simeq 0.34$\footnote{On inclusion of baryon acoustic oscillation (BAO) data the constraint becomes more stringent: $N_\text{eff} = 2.99 \pm 0.17$. A combined BBN+CMB analysis shows $N_\text{eff} = 2.880 \pm 0.144$~\cite{Yeh:2022heq}. Upcoming CMB experiments like CMB-S4~\cite{Abazajian:2019eic} and CMB-HD~\cite{CMB-HD:2022bsz} will be sensitive to a precision of $\DNeff \simeq 0.06$ and $\DNeff \simeq 0.027$, respectively. The next generation of satellite missions, such as COrE~\cite{COrE:2011bfs} and Euclid~\cite{EUCLID:2011zbd}, shall improve the limit even further up to $\DNeff \lesssim 0.013$.}. Now, the ratio of the gravitational wave (GW) energy density to that of the radiation bath is given by $\rho_{\rm GW}/\rho_R = \left(M_P^2/\rho_R\right)\,\left(k_{\rm hc}^2/8\right)\,\Delta^2_{h\,,{\rm inf}}(k)$~\cite{Boyle:2005se,Barman:2022qgt}, 
where $k_{\rm hc}=a_{\rm hc}\,H_{\rm hc}$ is the momentum mode, calculated at the moment it re-enters the horizon (``hc" denotes horizon-crossing). If horizon crossing occurs during radiation domination $k_{\rm hc}^2\propto\rho_R$, then the GW spectrum becomes scale invariant. On the other hand, if horizon crossing occurs during the inflaton-dominated era, the GW energy density is enhanced by a factor of $\rho_\phi / \rho_R$ evaluated at $T_{\rm hc}$. As a result, the largest enhancement occurs for the mode that re-enters the horizon right after inflation at $\Tmax$. For minimal gravitational reheating $(\xi_h = 0)$, such enhancement turns out to be $\rend / \rho_R\simeq (4-6) \times 10^{13}$ for $k\in[6,20]$, resulting in $\Omega_{\rm GW} h^2 \simeq (8-10) \times 10^{-6}$. This is  in clear conflict with the present bound from Planck on $\DNeff$, as discussed above. The constraint can be relaxed by increasing the value of $\Tmax$, since in that case the the energy density of GW relative to that of radiation at $\Tmax$ becomes smaller, compared to the minimal gravitational reheating scenario. The region labelled as `inconsistent reheating' in Fig.~\ref{fig:Trh} thus corresponds to $\xi_h<0$, which is forbidden from $\DNeff$ bound due to Planck on excessive GW production (shown in blue). 

Before closing this section, it is important to highlight for $k > 4$, the inflaton energy density redshifts faster than radiation, and the equation of state rapidly evolves from some stiffer fluid $w_\phi > 1/3$ to $w_\phi = 1/3$, and the universe is dominated by massless inflaton particles (not a condensate anymore), with energy density redshifting as $a^{-4}$~\cite{Lozanov:2016hid,Lozanov:2017hjm}. These free particles are produced as a consequence of the self-interaction of the inflaton, a process known as {\it fragmentation}. A detailed study of such processes, taking into account the effect of parametric and tachyonic resonance, requires dedicated lattice simulations and has been studied, for example, in  Refs.~\cite{Greene:1997fu,Green:1999yh,Amin:2011hj, Lozanov:2016hid,Figueroa:2016wxr,Lozanov:2017hjm,Garcia:2023eol,Garcia:2023dyf}. This, however, is beyond the scope of the present draft. 

\section{Axions confronting gravitational reheating}
\label{sec:qcd-axion}
Having discussed the details of the gravitational reheating scenario, we will now move on to the discussion of producing QCD axions (and axion like particles) via standard misalignment mechanism {\it during} the epoch of reheating. Our main focus is to explore the relevant parameter space where all of the observed DM abundance is produced by axions, when misalignment happens during the era of inflaton domination. Before moving on, we would first like to give a brief review of the axion model, mentioning some of the relevant quantities to take into account.  

The Lagrangian density for axion field  reads~\cite{GrillidiCortona:2015jxo}
\begin{align}
\mathcal{L}_a \supset \frac{1}{2}\partial^\mu a\, \partial_\mu a - \tilde{m}^2_a(T) f_a^2 \left[1-\cos\left(\frac{a}{f_a}\right)\right]\,,
\end{align}
which leads to equation of motion of the zero modes as
\begin{align}\label{eq:thetaeq}
\ddot{\theta} + 3 H\, \dot{\theta} + \tilde{m}^2_a(T)\, \sin \theta =0\,,
\end{align}
where  $\theta(t) \equiv a(t)/f_a$ and $f_a$ denotes the decay constant.  Here, the temperature dependent mass of the axion is denoted by $\tilde{m}_a(T)$, which for the QCD axion, is shown to be dependent on the topological susceptibility of QCD $\chi(T)$ via~\cite{Borsanyi:2016ksw}
\begin{align}
& \tilde{m}_a(T) = \sqrt{\chi(T)}/f_a\,. 
\end{align}
The lattice QCD simulations have provided an estimates of the zero-temperature value of $\chi(T)$, which is given by $\chi_0\equiv\chi(0)\simeq 0.0245\,\text{fm}^{-4}$, in the symmetric isospin case. Subsequently, the form of the $\tilde{m}_a(T)$ is found to be
\begin{equation}\label{eq:axion-mass}
    \tilde m_a(T) \simeq m_a \times
    \begin{cases}
	    (\Tqcd/T)^4 & \text{for } T \geq \Tqcd\,,\\
	    1 & \text{for } T \leq \Tqcd\,,
    \end{cases}
\end{equation}
with  $m_a$ representing the axion mass at zero temperature and is given by~\cite{DiLuzio:2020wdo}
\begin{equation}
    m_a \simeq 5.7 \times 10^{-6} \left(\frac{10^{12}~\text{GeV}}{f_a}\right) \text{eV}\,.
\end{equation}

Following Eq.~\eqref{eq:thetaeq}, axion begins to oscillate at the temperature $T = \Tosc$ defined by $3\,H(\Tosc) \equiv \tilde m_a(\Tosc)$~\cite{Kolb:1990vq}.  Assuming a radiation dominated Universe, the corresponding oscillation temperature can be evaluated as
\begin{align}\label{eq:Tosc_RD}
\Tosc \simeq 
\begin{cases} 
\left(\frac{1}{\pi}\sqrt{10/\gs(\Tosc)}\,m_a\, M_P \right)^{1/2} &  \Tosc \leq \Tqcd \,,\\
\left(\frac{1}{\pi} \sqrt{10/\gs(\Tosc)}\,m_a\, M_P\, \Tqcd^4\right)^{1/6}   & \Tosc \geq  \Tqcd,
\end{cases}
\end{align}
where the Hubble expansion rate takes the form $H(T) = \sqrt{\rR(T)/(3\,M_P^2)}$. Below $\Tosc$, the axion can be considered as a non-relativistic particle. Considering the conservation of the axion number density and assuming conservation of SM entropy, the energy density for such non-relativistic axions $\rho_a$ at present is given by
\begin{equation} \label{eq:rho0}
    \rho_a(T_0) = \rho_a(\Tosc) \frac{m_a}{\tilde m_a(\Tosc)} \frac{s(T_0)}{s(\Tosc)}\,,
\end{equation}
with $T_0$ representing the temperature today and the SM entropy density is defined as 
\begin{equation}
    s(T) = \frac{2\pi^2}{45}\, \gss(T)\, T^3\,,
\end{equation}
where $\gss(T)$ denotes the corresponding  number of relativistic degrees~\cite{Drees:2015exa}. The
WKB approximation leads to $\rho_a(\Tosc) \simeq \frac12 \tilde m_a^2(\Tosc)\, f_a^2\, \theta_i^2$, where $\theta_i$ is the initial misalignment angle~\cite{Hertzberg:2008wr, DiLuzio:2020wdo}. 
Eventually, using Eq.~\eqref{eq:rho0}, the axion abundance can be found to be
\begin{equation}
    \begin{aligned}
        \Omega_a & \equiv \frac{\rho_a(T_0)}{\rho_c} \simeq \left(\frac{\theta_i}{1}\right)^2\times
        \begin{cases}
          0.002\times\left(\frac{m_a}{5.6\,\mu{\rm eV}}\right)^{-3/2} & \text{for } m_a \lesssim m_a^\text{qcd},\\
            0.09\times\left(\frac{m_a}{5.6\,\mu{\rm eV}}\right)^{-7/6} & \text{for } m_a \gtrsim m_a^\text{qcd},
        \end{cases}
    \end{aligned}
\end{equation}
with $m_a^\text{qcd} \equiv m_a(\Tosc=\Tqcd) \simeq 4.8 \times 10^{-11}$~eV and $g_\star(\Tqcd)\simeq 25$. Here $\rho_c/h^2 \simeq 1.1 \times 10^{-5}$~GeV/cm$^3$ is the critical energy density and $s(T_0) \simeq 2.69 \times 10^3$~cm$^{-3}$ is the entropy density at present~\cite{Planck:2018vyg}.

Next, we take up the scenario when the oscillation of the axion starts during the reheating period i.e., $\Tosc>\Trh$. In this case, the axion energy density at present epoch reads
\begin{align}\label{eq:rhoaRH}
& \rho_a(T_0)=\rho_a(\Tosc)\,\frac{m_a}{\tilde{m}_a(\Tosc)}\,\frac{s(T_0)}{s(\Tosc)}\times\frac{S(\Tosc)}{S(\Trh)}\,,
\end{align}
where the last factor takes into account the dilution of the axion energy density due to the entropy injection as a result of energy transfer from the inflaton to the bath. This dilution factor can be determined as
\begin{align}
&\frac{S(T)}{S(\Trh)} =   \frac{\gss(T)}{\gss(\Trh)}\,\left(\frac{T}{\Trh}\right)^3\,\left(\frac{a(T)}{\arh}\right)^3 = \frac{\gss(T)}{\gss(\Trh)}\,,   
\end{align}
where we have used Eq.~\eqref{eq:rhoR}. Note that since the non trivial expansion of the Universe is the sole reason for the dilution of radiation, the factor accounting for entropy injection is simply ratio of DOFs, unlike the case in~\cite{Xu:2023lxw}. Now, one can further consider two sub-cases depending on the hierarchy between $\Tqcd$ and $\Tosc$. Below we discuss them one by one.
\begin{figure}[t!]
    \centering
    \includegraphics[scale=0.34]{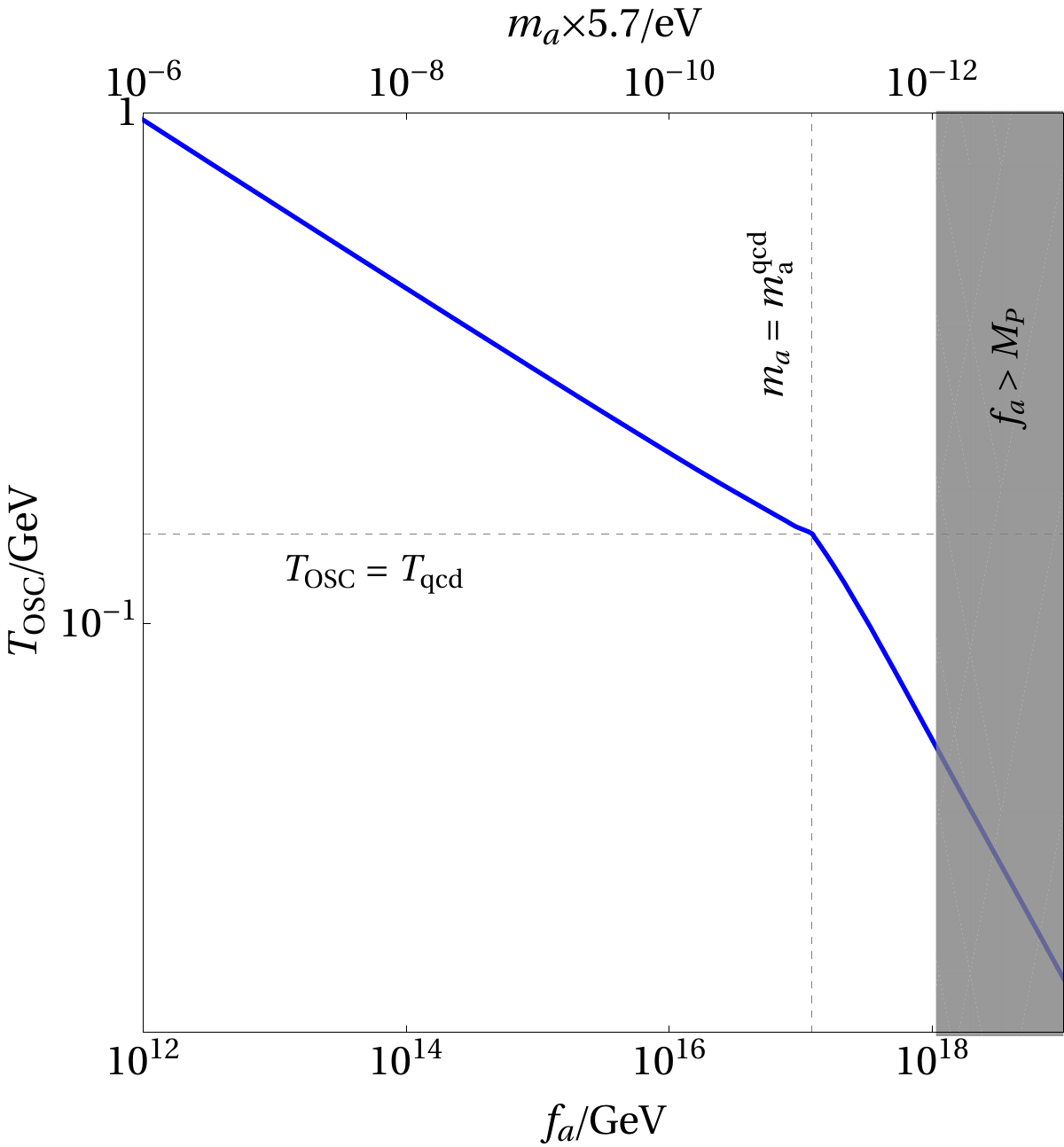}~~~~\includegraphics[scale=0.34]{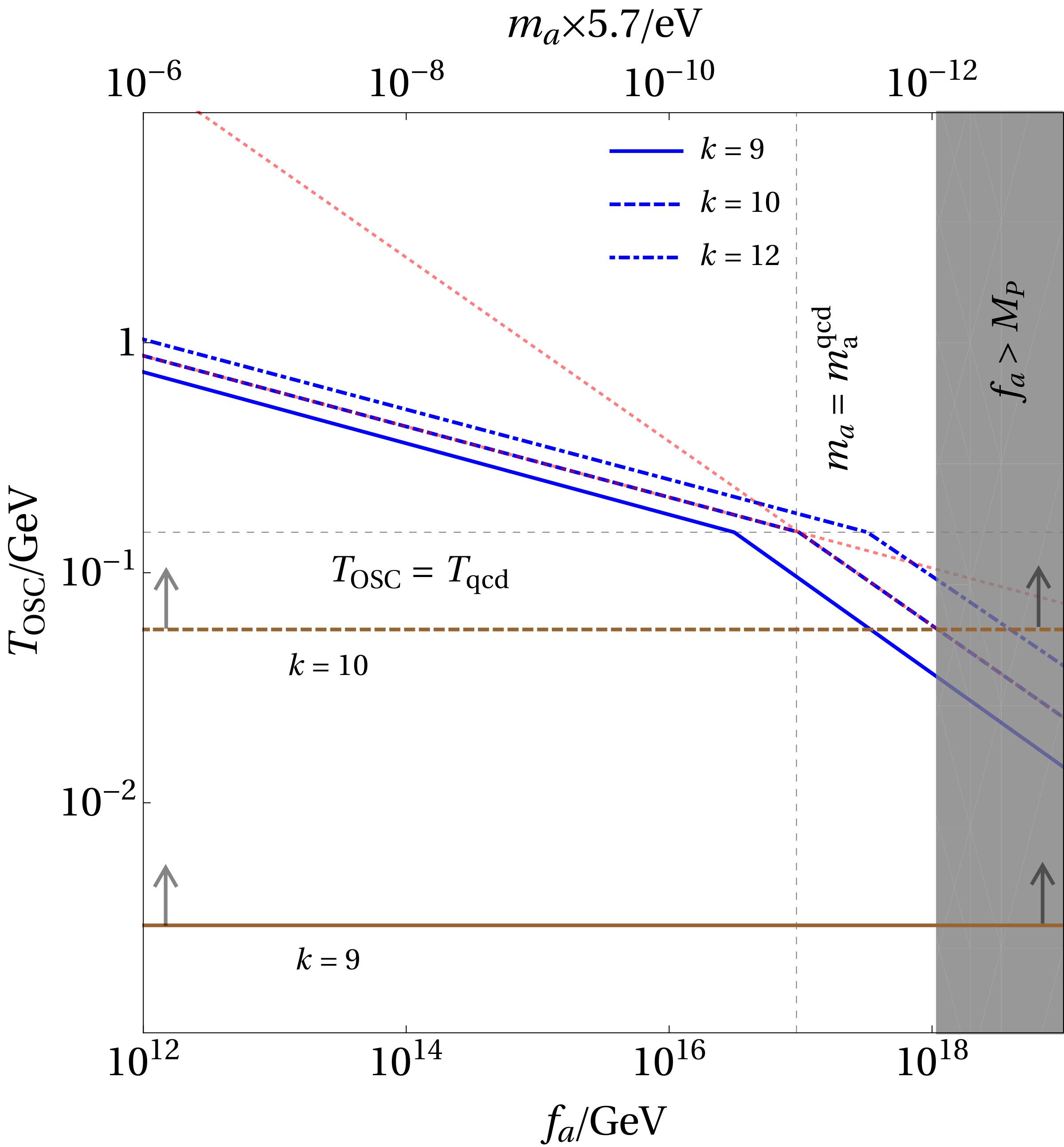}
    \caption{{\it Left:} QCD axion oscillation temperature as a function of axion mass in radiation dominated background. {\it Right:} Same as left, but in an inflaton dominated background for different choices of $k~(\equiv \Trh)$, considering {\it minimal} gravitational reheating. The red dotted lines correspond to analytical solutions [cf. Eq.~\eqref{eq:Tosc1} and \eqref{eq:Tosc2}]. The brown horizontal lines correspond to $\Tosc=\Trh$ for $k=\{9,\,10\}$ from bottom to top. For each case the arrowheads show the region of parameter space where oscillation happens during reheating.}
    \label{fig:Tosc}
\end{figure} 

\subsubsection*{Scenario-I:}
We first consider the case $\Tqcd<\Tosc$, for which the axion mass shows a temperature dependence $\tilde{m}_a=m_a\,\left(T/\Tqcd\right)^{-4}$. One can then obtain the expression for oscillation temperature as
\begin{align}\label{eq:Tosc1}
& \Tosc = \Trh\,\left(\frac{1}{\pi}\,\sqrt{\frac{10}{\gs(\Trh)}}\,\frac{m_a\,M_P\,\Tqcd^4}{\Trh^6}\right)^\frac{2+k}{8+7\,k}\,.    
\end{align}
Note that, this scenario can be further classified as: $\Trh<\Tqcd<\Tosc$ and $\Tqcd<\Trh<\Tosc$. The first inequality provides
\begin{align}
&  \Tqcd^\frac{3\,k}{k-4}\,\left(\frac{1}{\pi}\,\sqrt{\frac{10}{\gs(\Trh)}}\,m_a\,M_P\right)^\frac{2+k}{4-k}<\Trh<\Tqcd\,,   
\end{align}
while from the second inequality we get
\begin{align}
& \Tqcd<\Trh<\left(\frac{1}{\pi}\,\sqrt{\frac{10}{\gs(\Trh)}}\,m_a\,M_P\,\Tqcd^4\right)^{1/6}\,.
\end{align}

\subsubsection*{Scenario-II:}
\begin{figure}[t!]
    \centering
    \includegraphics[scale=0.6]{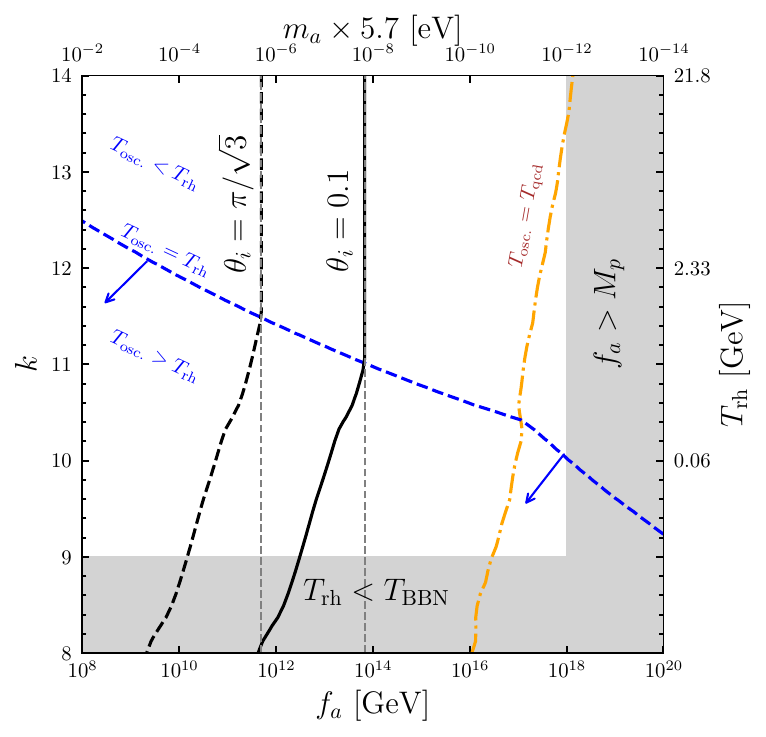}~~~~
    \includegraphics[scale=0.48]{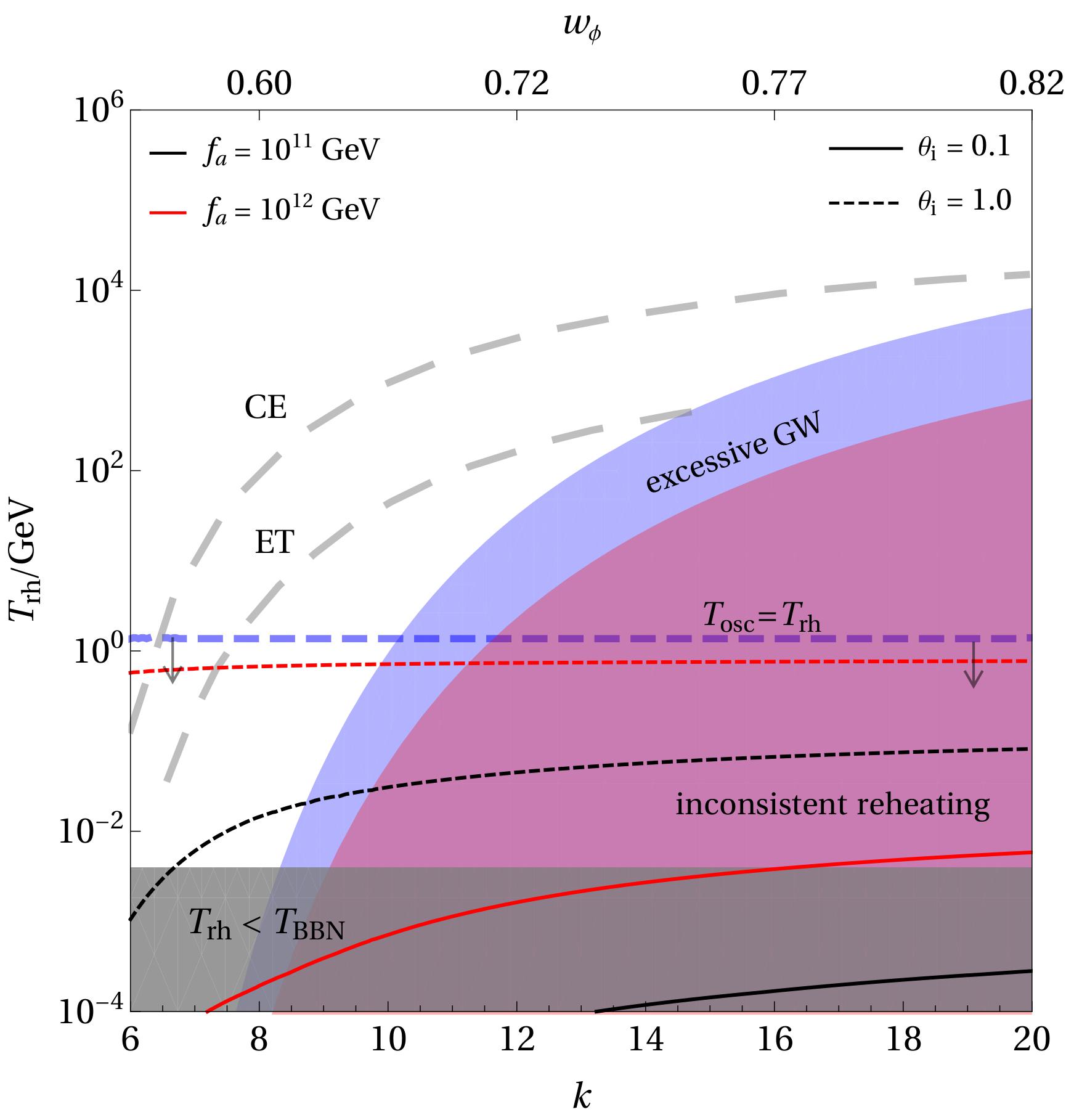}
    \caption{{\it Left:} The black curves satisfy the observed relic abundance for $\theta_i=0.1$ (thick) and $\theta_i=\pi/\sqrt{3}$ (dashed), considering {\it minimal} gravitational reheating. The gray shaded regions are disallowed from BBN bound on $\Trh\,(k\geq 9)$ and super-Planckian scale $f_a>M_P$. Along each vertical gray dashed line total relic abundance is produced if oscillation takes place during radiation domination. {\it Right:} Contours producing observed relic abundance, in case when the radiation has non-minimal contribution to radiation for $f_a=\{10^{11}\,,10^{12}\}$ GeV depicted by black and red contours. The solid and dashed patterns represent $\theta_i=0.1$ and $\theta_i=1$ respectively. The red shaded region is ruled out from BBN bound on $\DNeff$ for overproduction of GW. The gray shaded region is ruled out from BBN bound on $\Trh\gtrsim 4$ MeV. The gray dashed curves correspond to exclusion limits from future GW detectors. In both cases the region of parameter space of our interest $(\Tosc>\Trh)$ is shown by arrowheads.}
    \label{fig:relic}
\end{figure}

In the second case, $\Tqcd>\Tosc$, the axion mass remains constant. In this case we can again derive an analytical expression for the oscillation temperature as
\begin{align}\label{eq:Tosc2}
& \Tosc =  \Trh\,\left(\frac{1}{\pi}\,\sqrt{\frac{10}{\gs(\Trh)}}\,\frac{m_a\,M_P}{\Trh^2}\right)^\frac{k+2}{3\,k}\,,   
\end{align}
for $\tilde {m}_a=m_a$. Using the inequality $\Trh<\Tosc$ one finds
\begin{align}
& \Trh < \left(\frac{10}{\pi^2\,\gs(\Trh)}\right)^\frac{1}{4} \,\sqrt{m_a\,M_P}\,,
\end{align}
while $\Tosc<\Tqcd$ further provides
\begin{align}
&  \Trh < \Tqcd^\frac{3\,k}{k-4}\,\left(\sqrt{\frac{10}{\gs(\Trh)}}\,\frac{m_a\,M_P}{\pi}\right)^{-\frac{k+2}{k-4}}\,.   
\end{align}
In Fig.~\ref{fig:Tosc} we show QCD axion oscillation temperature as a function of the scale $f_a$. In the left panel we consider standard radiation dominated Universe, while in the right panel we consider an inflaton-dominated background. In either cases we see a bend around $\Tosc\simeq\Tqcd$ due to the change in the temperature dependence of the axion mass. In the right panel we note, depending on the choice of $k$ (that decides steepness of the inflaton potential) the oscillation temperature changes, following Eq.~\eqref{eq:Tosc1} and \eqref{eq:Tosc2}. The analytically derived expressions for $\Tosc$ are denoted with the red dotted lines for one of the values of $k$, showing the agreement between analytical and numerical results. The gray shaded region in both plots is forbidden as it surpasses the Planck scale. While $k>9$ is necessary to have successful BBN in case of minimal gravitational reheating, as $k$ grows, $\Trh$ starts increasing till it reaches $k\simeq 12$, above which $\Trh>\Tosc$, rendering the oscillation to take place during radiation domination. So, in case of minimal gravitational reheating our parameter space is confined within $9<k\lesssim 12$ (this is more prominent from the right panel Fig.~\ref{fig:relic} which will be discussed in a moment), for which misalignment happens during reheating. In case of gravitational reheating via non-minimal coupling, one can, however, go to higher $k$-values. 

Utilizing Eq.~\eqref{eq:rhoaRH}, we can obtain the relic abundance for $\Trh<\Tosc$ as
\begin{align}\label{eq:rel-rh}
& \Omega_a\,h^2 = \left(\frac{\theta_i}{1}\right)^2
\begin{cases}
571.2\times\left(\frac{m_a}{5.6\,\mueV}\right)^{-6/13}\,\left(\frac{\Trh}{T_{\rm BBN}}\right)^{-15/71}  & \text{for } m_a \lesssim m_a^\text{qcd}\\[8pt]
1.2\times 10^{-7}\times\left(\frac{m_a}{5.6\,\mueV}\right)^{-14/5}\,\left(\frac{\Trh}{T_{\rm BBN}}\right)^{-7/5}   & \text{for } m_a \gtrsim m_a^\text{qcd}\,,
\end{cases}
\end{align}
where we have fixed $k=10$, which also fixes $\Trh=0.05$ GeV in case of minimal reheating scenario. In the left panel of Fig.~\ref{fig:relic} we show parameter space in $[k,\,f_a]$ plane that reproduces all of the Planck observed relic abundance (shown via black curve), considering misalignment happens during gravitational reheating. As we see, a larger initial misalignment angle shifts the parameter space to lower $f_a$ (or heavier axion mass) since the relic abundance varies inversely with $m_a$ for a fixed $\theta_i$, as seen from Eq.~\eqref{eq:rel-rh}. We show contours corresponding to $\Tosc=\Trh$ and $\Tosc=\Tqcd$ via blue and orange broken curves respectively. As a result, parameter space of our interest $(\Tosc>\Trh)$ lies below the blue dashed contour. The shaded regions are disallowed from $\Trh<\TBBN$ (or, equivalently, $k\lesssim 9$) and $f_a>M_P$. For each $k$, we denote corresponding $\Trh$ (fixed for $\xi_h=0$) along the right vertical axis. The gray vertical dashed lines show required $f_a$ that can produce the observed relic abundance if the oscillation happens during radiation dominated background. For the same axion mass, gravitational reheating thus opens up larger parameter space, satisfying DM abundance. More precisely, for $\theta_i=\pi/\sqrt{3}$, misalignment during RD produces right abundance for $f_a\simeq 10^{12}$ GeV (equivalently, $m_a\simeq 4.2\times 10^{-6}$ eV), whereas for minimal gravitational reheating this window becomes $10^{10}\lesssim f_a\lesssim 10^{12}$ GeV (or, $10^{-4}\lesssim m_a\lesssim 10^{-6}$ eV). In the right panel of Fig.~\ref{fig:relic}, the red and black curves correspond to contours producing observed DM abundance for different choices of $\theta_i$, where we fix $f_a=\{10^{11},\,10^{12}\}$ GeV. Here we consider {\it non-minimal} contribution to the radiation, and show the resulting parameter space in $[\Trh-k]$ plane. The relic density satisfying contours correspond to $\theta_i=\{0.1,\,1.0\}$, shown via solid and dashed pattern respectively. As mentioned in Sec.~\ref{sec:model}, for each $k$, in case of non-minimal gravitational reheating, one now have the freedom choose appropriate $\xi_h$ that can provide the desired $\Trh$ (larger than that corresponding to minimal reheating scenario). The region below the blue  dashed line, parallel to the horizontal axis, indicates $\Tosc>\Trh$. Here we see, a larger $\theta_i$ requires larger $\Trh$, as one can infer from Eq.~\eqref{eq:rel-rh}. Also, for a fixed $\theta_i$, lower $f_a$ (higher $m_a$) requires lower $\Trh$ in order to produce right abundance, again following Eq.~\eqref{eq:rel-rh}. As one can notice, the $\DNeff$ bound plays a very crucial role in constraining the viable parameter space, ruling out $\theta_i\lesssim 0.1$ for both $f_a$'s. 
\begin{figure}[t!]
    \centering
    \includegraphics[scale=0.6]{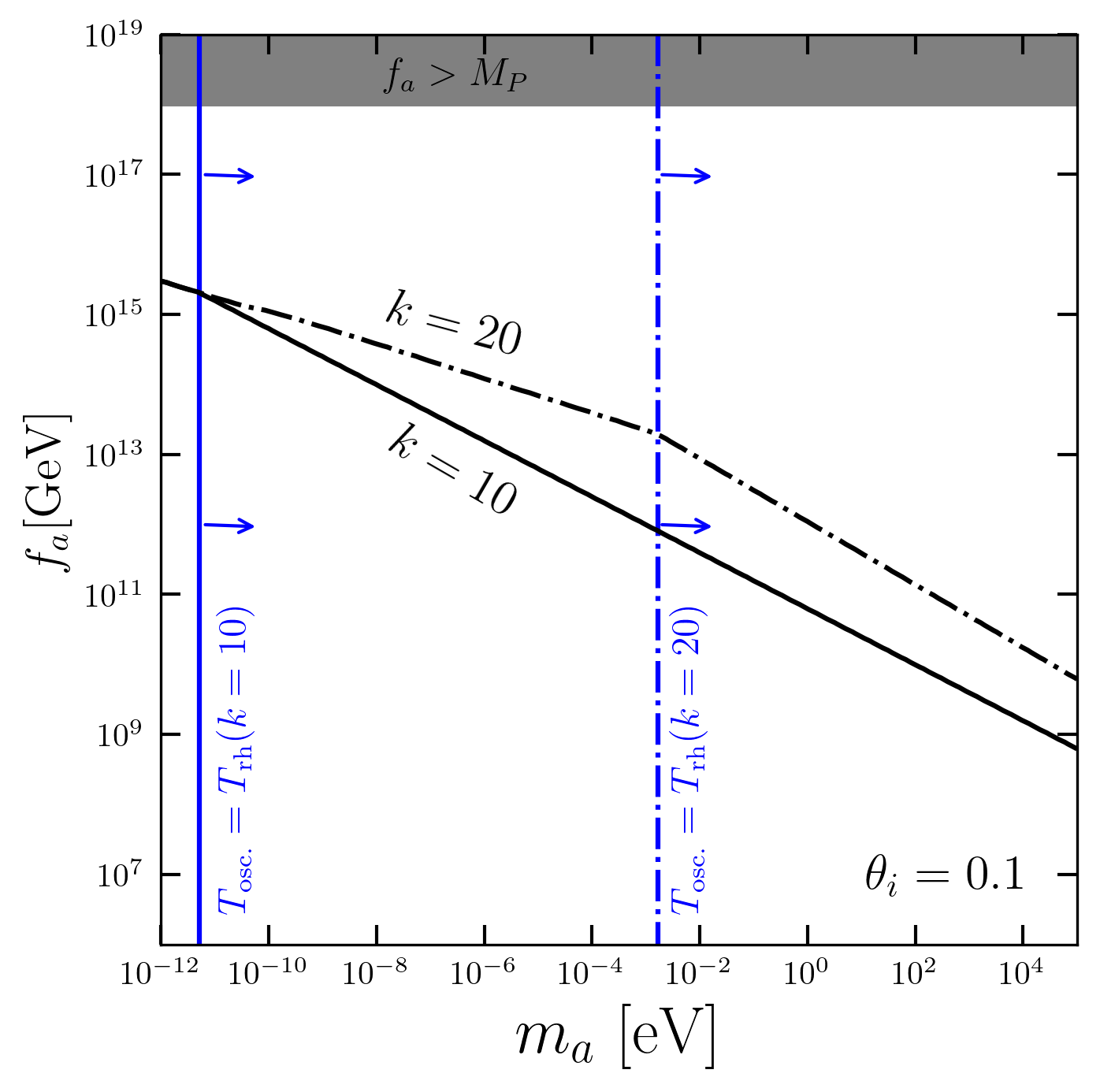}~~~~\includegraphics[scale=0.6]{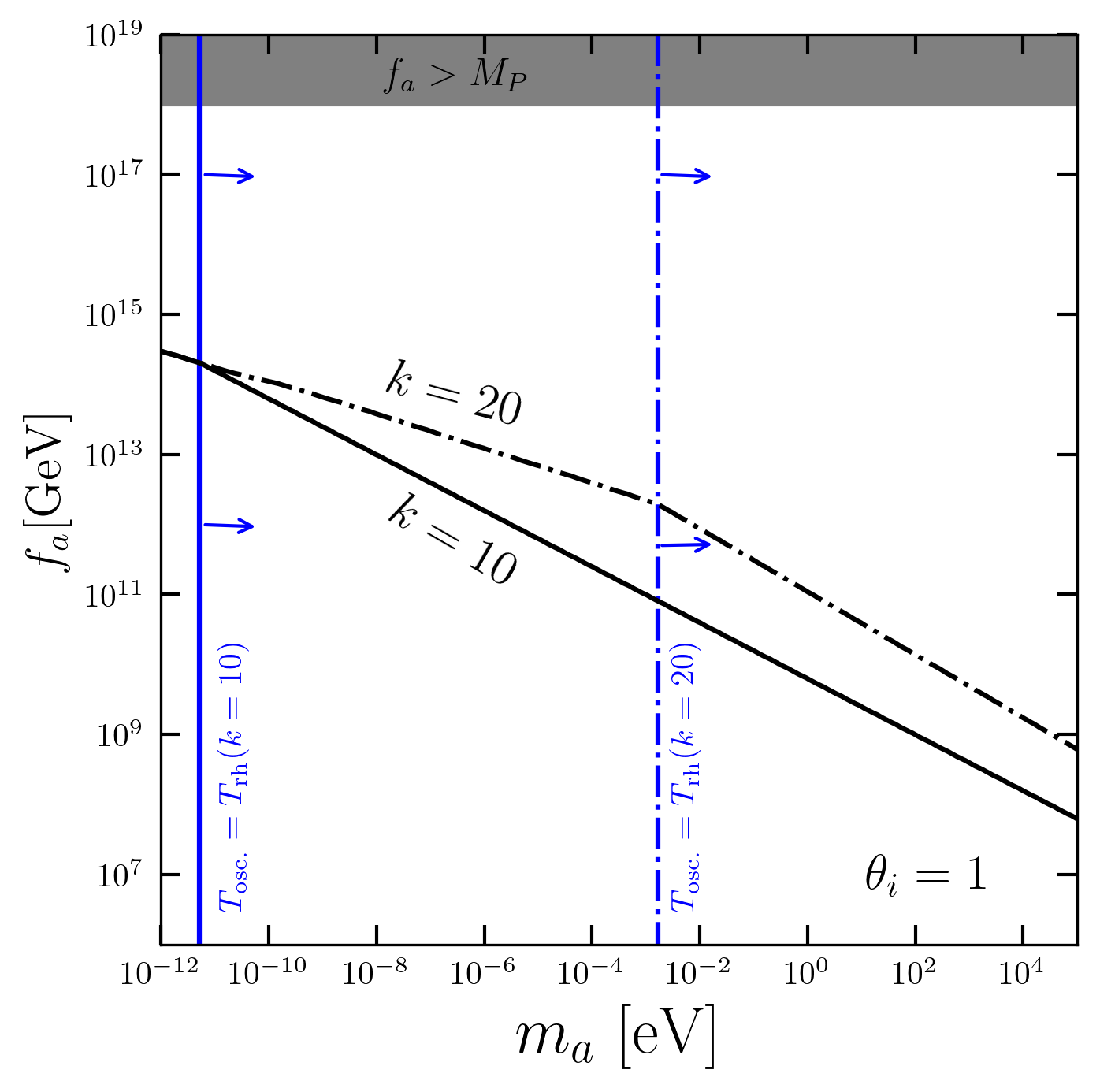}\\[10pt]
    \includegraphics[scale=0.36]{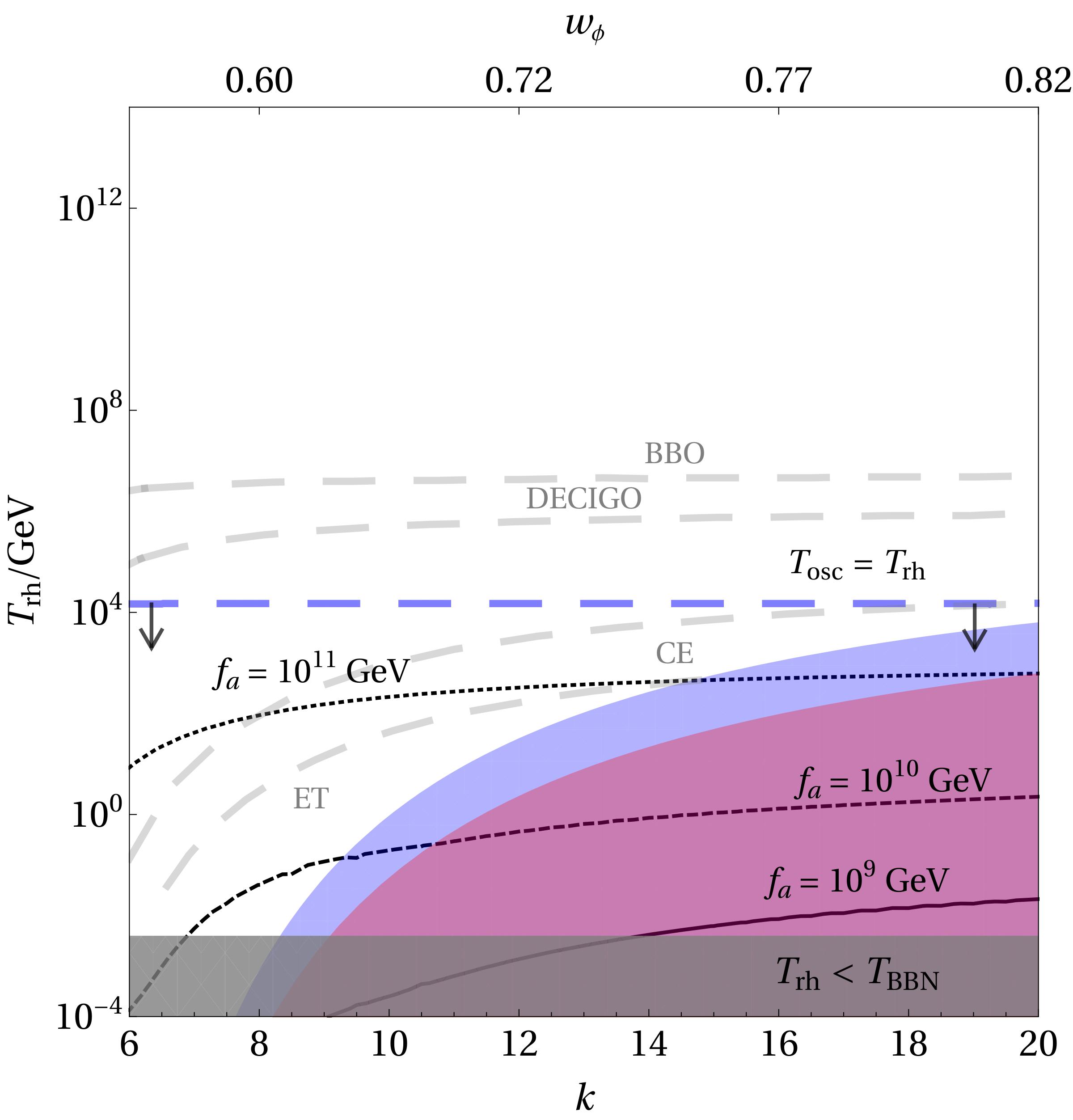}~~~~\includegraphics[scale=0.36]{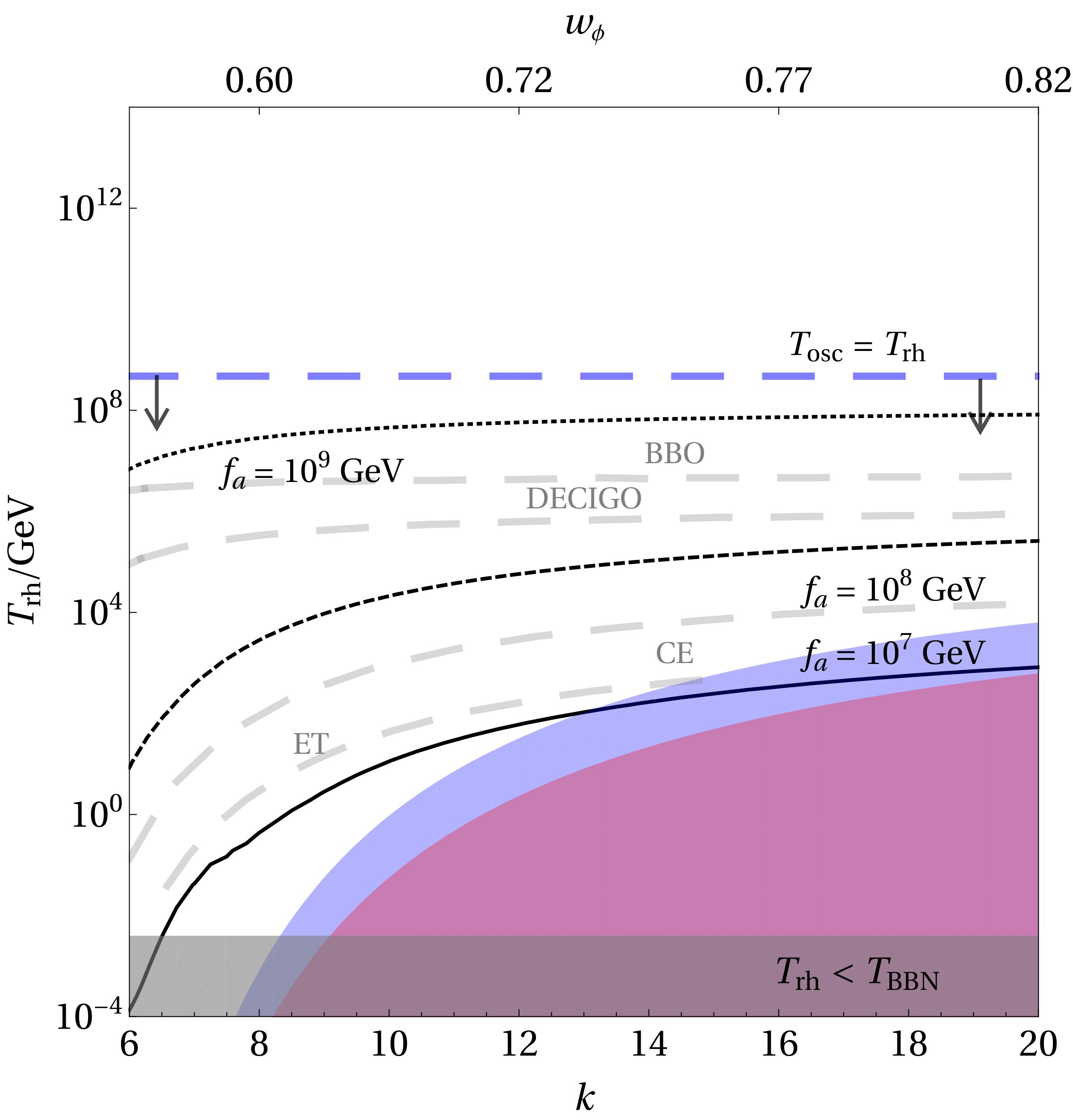}
    \caption{{\it Top:} The black contours satisfy the observed relic abundance in case of ALPs for different choices of $k$ shown by different patterns for the {\it minimal} reheating scenario. Here we choose $\theta_i=\{0.1,\,1\}$ in the left and in the right panel respectively. The blue vertical straight lines correspond to $\Tosc=\Trh$ for each $k$. The gray shaded region is discarded by super-Planckian bound on $f_a$. {\it Bottom:} Contours producing observed relic abundance for $\theta_i=1$, $m_a=1$ eV (left panel) and $m_a=1$ GeV (right panel), for different choices of $f_a$ shown with different patterns. Other constraints are same as those appearing in Fig.~\ref{fig:relic}. In all cases the region of parameter space of our interest $(\Tosc>\Trh)$ is shown by arrowheads.}
    \label{fig:relALP}
\end{figure} 

Finally, we turn our attention towards ALPs, for which we consider the mass $m_a$ to be time independent. For oscillations during reheating $\Trh<\Tosc$ we find the oscillation temperature to be
\begin{align}\label{eq:Tosc-ALP}
& \Tosc = \Trh\,\left[\frac{m_a\,M_P}{\Trh^2}\,\frac{1}{\pi}\,\sqrt{\frac{10}{\gs(\Trh)}}\right]^\frac{k+2}{3\,k}\,,    
\end{align}
same as Eq.~\eqref{eq:Tosc2}. This puts a bound on the reheating temperature, which in turn can be translated into a lower bound on the ALP mass via 
\begin{align}
& T_{\rm BBN}\lesssim\Trh\lesssim\Tosc
\implies
m_a \gtrsim \sqrt{\frac{\pi^2\,\gs}{10}}\,\frac{T_{\rm BBN}^2}{M_P}\,.
\end{align}
The ALP energy density at present day reads
\begin{align}\label{eq:rho-ALP}
& \rho_a(T_0)=\rho_a(\Tosc)\,\frac{s(T_0)}{s(\Tosc)}\,\frac{\gs(\Tosc)}{\gs(\Trh)}\,, 
\end{align}
with $\rho_a(\Tosc)\simeq (1/2)\,m_a^2\,f_a^2\,\theta_i^2$, this turns out to be
\begin{align}\label{eq:rho-ALP2}
& \frac{\rho_a}{s}\Big|_{T_0}
\simeq\frac{45\,\pi^{2/k}}{4\,\pi}\,\frac{\gs(\Tosc)}{\gs(\Trh)\,\gss(\Tosc)}\,\left(\sqrt{\frac{10}{\gs(\Trh)}}\frac{m_a\,M_P}{\Trh^2}\right)^{-\frac{2+k}{k}}\,\theta_i^2\,\frac{f_a^2\,m_a^2}{\Trh^3}
\nonumber\\&
\xrightarrow[]{k\gg 2}
\frac{45}{4}\,\frac{\gs(\Tosc)}{\gs(\Trh)\,\gss(\Tosc)}\,\,\sqrt{\frac{\pi^2\,\gs(\Trh)}{10}}\,\theta_i^2\,
\frac{f_a^2\,m_a}{\Trh\,M_P}\,.    
\end{align}
In case of ALPs, $f_a$ and $m_a$ can vary independently, that makes following set of parameters free: $\{m_a,\,f_a,\,\theta_i,\,k\}$ for minimal and $\{m_a,\,f_a,\,\theta_i,\,k,\,\Trh\}$ for non-minimal reheating scenarios. From Eq.~\eqref{eq:rho-ALP} we find, in order to satisfy the observed DM abundance, $f_a^2\propto m_a^\frac{2-k}{k}$, implying, a larger $f_a$ requires lighter ALP for a fixed $k>2$ (and $\theta_i$). This is what we see in the top panel of Fig.~\ref{fig:relALP}, where the black thick contours satisfy the observed relic abundance for different choices of $k$ with fixed $\theta_i$'s. Corresponding to each $k$-values, we also show contours satisfying $\Tosc=\Trh$ in blue. Since in the minimal reheating scenario fixing $k$ fixes the reheating temperature, hence we see simple straight line contours parallel to the $f_a$ axes. The change in slope for each contour occurs at $\Tosc=\Trh$, indicated by the blue vertical straight lines. To the left of each vertical line oscillation occurs during RD, while to the right during reheating. The parameter space of our interest therefore lies to the right of each blue vertical line. In the bottom panel of Fig.~\ref{fig:relALP} we show contours of correct relic density for different choices of $f_a$, while fixing $m_a$ for a given $\theta_i=1$, now considering non-minimal gravitational reheating. Again, the allowed parameter space lies below the blue dashed curve, for which $\Tosc=\Trh$. As we see, for $m_a=1$ eV (bottom left panel), $f_a\lesssim 10^9$ GeV is completely ruled out from BBN bound on $\Trh$, together with $\DNeff$ constraint on GW energy density, irrespective of the choice of $k$. This can be relaxed by considering a heavier $m_a=1$ GeV, as seen from the bottom right panel. The reason being, the relic density iso-contours satisfy $f_a^2\,m_a\propto\Trh$ for a fixed $\theta_i$, following Eq.~\eqref{eq:rho-ALP2}. Therefore, for the same choice of $f_a$ and $\theta_i$, heavier ALPs require higher $\Trh$ to produce the right abundance. 

\section{Purely gravitational axions}
\label{sec:grav-axion}
To this end we have discussed axion production via standard misalignment mechanism in an inflaton-dominated background. However, due to the inevitable gravitational interactions, through Eq.~\eqref{eq:grav-lgrng}, axions are also produced via gravity-mediated 2-to-2 process, which we refer to as {\it purely gravitational production} of axions. Such purely gravitational production can happen from (i) thermal bath and (ii) scattering of the inflaton condensate. Below we discuss the details of each such processes and obtain the resulting parameter space.

\subsection{Thermal scattering}
\label{sec:thermal}
In case of production from thermal bath due to scattering of the SM particles mediated by gravity, the rate of production reads~\cite{Bernal:2018qlk,Clery:2022wib}
\begin{align}
& R_a^T = \frac{3997\,\pi^3}{41472000}\,\frac{T^8}{M_P^4}\equiv\beta_0\,\frac{T^8}{M_P^4}\,.    
\end{align}
Since we are interested in axion production during reheating, it is instructive to consider the comoving axion number $N_a=n_a\times a^3$ as the entropy is not conserved during reheating. To track the comoving axion number, we solve the corresponding Boltzmann equation   
\begin{align}
& \frac{dN_a}{da}=\frac{a^2\,R_a^T}{H}\,.  
\end{align}
One can then find the axion number density $n_a$ at the end of reheating as~\cite{Barman:2022qgt}
\begin{align}\label{eq:nT}
& n^T_a(\arh)=\frac{\beta_0\,(k+2)\,\rRH^\frac{3}{2}}{12\,\sqrt{3}M_P^3\,\alpha_*^2}
\left(\frac{1}{1-{\tt {a}}^{\frac{14-8k}{k+2}}}\right)^2\,
\nonumber\\ &
\left[\frac{2\,(7-4k)^2}{(k+5)\,(k-1)\,(5k-2)}\,{\tt {a}}^{\frac{10+2k}{k+2}}-\frac{9}{(k+5)} + \frac{18}{(5k-2)}\,{\tt {a}}^{\frac{14-8k}{k+2}} - \frac{1}{(k-1)}\, {\tt {a}}^{\frac{28-16k}{k+2}}\right]\,,
\end{align}
where we have made use of Eq.~\eqref{eq:rhoR}. Since the gravitational reheating temperature is generally quite low [cf. Sec.~\ref{sec:grav-reheat}], considering ${\tt {a}}\gg 1$ we obtain
\begin{equation}
    n^T_a (\arh) \simeq \frac{\beta_0\,(k+2)\,\rRH^\frac{3}{2}}{12\,\sqrt{3}\,M_P^3\,\alpha_*^2} \frac{2\,(7-4k)^2}{(k+5)\,(k-1)\,(5k-2)}\,{\tt {a}}^{\frac{10+2k}{k+2}}\,,
    \label{eq:nT2}
\end{equation}
where ${\tt{a}}=\arh/\aend\equiv \left(H_{\rm end}/\Hrh\right)^{k+2/(3\,k)}$ [cf. Eq.~\eqref{eq:hubble}] and $\alpha_\star=\left(\pi^2/30\right)\,\gs(\Trh)$. 
\subsection{Inflaton scattering}
\label{sec:inflaton}
For gravitational production from inflaton condensate scattering, {\it viz.,} $\phi\phi\to aa$, the particle production rate per unit volume and unit time for arbitrary $k$ reads~\cite{Clery:2021bwz,Clery:2022wib,Barman:2022qgt}
\begin{align}
& R_a^{\phi^k}=\frac{2\,\rho_\phi^2}{16\,\pi\,M_P^4}\,\Sigma_0^k\,,    
\end{align}
where
\begin{align}
& \Sigma_0^k = \sum_{n=1}^\infty\,|\mathcal{P}_n^k|^2\,\left(1+\frac{2\,m_a^2}{E_n^2}\right)^2\,\sqrt{1-\frac{4\,m_a^2}{E_n^2}}\,.    
\end{align}
The factor of 2 accounts for pair production. Here $E_n=n\,\omega$ is the energy of the $n$-th inflaton oscillation mode. Note that, such a production is possible only during reheating and not after, unlike the production from radiation bath that can take place both during and post-reheating. The evolution of comoving axion number produced from inflaton scattering mediated by graviton reads
\begin{align}
& \frac{dN_a}{da}=\frac{a^2\,R_a^{\phi^k}}{H}\,.
\end{align}
Again, the number density of axions at the end of reheating can be computed as
\begin{align}\label{eq:nphi}
& n_a^\phi(\arh)=\frac{\sqrt{3}\,\rRH^{3/2}}{8\,\pi\,M_P^3}\,\frac{k+2}{6\,k-6}\,\left(\frac{\rend}{\rRH}\right)^{1-\frac{1}{k}}\,\Sigma_0^k\,.    
\end{align}
Taking both contributions into account, we determine the gravitationally produced axion relic abundance as
\begin{align}\label{eq:rel-grav}
& \Omega_a\,h^2\equiv\Omega_a^T\,h^2+\Omega_a^\phi\,h^2=1.6\times 10^8\,\frac{\gs(T_0)}{\gs(\Trh)}\,\frac{n_a(\Trh)}{\Trh^3}\,\frac{\matld(\Trh)}{1\,{\rm GeV}}\,,   
\end{align}
where $n_a(\Trh)=\sum_{j=\{\phi,\,T\}}n_a^j(\Trh)$. It is interesting to note, the ratio of  axion number density produced from thermal and from inflaton scattering reads
\begin{align}
& \frac{n_a^T}{n_a^{\phi^k}}\simeq\frac{\beta_0\,\pi}{\alpha_\star^2\,\Sigma_0^k}\,\frac{8\,(7-4\,k)^2}{3\,(5+k)\,(5\,k-2)}\,\left(\frac{\rho_{\rm rh}}{\rho_{\rm end}}\right)^\frac{2\,k-8}{3\,k}\ll 1\,,  
\end{align}
since $\rho_{\rm end}\gg\rRH$. This implies that purely gravitational production of axions is always dominated by the scattering of the inflaton zero modes. 

Now, axions produced from inflaton scattering during reheating are relativistic as the average momenta they carry is of the order of the inflaton mass, which is $\sim 10^{13}$ GeV at the beginning of reheating. Such a (semi-)relativistic population of axions at CMB decoupling behave as dark radiation and contribute to $\DNeff$. Following the analysis in Refs.~\cite{Acero:2008rh,Arias:2023wyg},the average thermal velocity of axions today $\langle v_{a,0}\rangle$ is related to $\DNeff$ and relic abundance $\langle v_{a,0}\rangle \simeq 5.62\times 10^{-6}\times\left(\DNeff/\Omega_a\,h^2\right)\,,$ where we assume the axions are relativistic/semi-relativistic at photon decoupling, a legitimate assumption for masses around the eV scale. Such a population of (semi-) relativistic axion is distinguishable from CDM if $\langle v_{a,0}\rangle\gtrsim$ 1 km/s. 
\begin{figure}[t!]
    \centering
    \includegraphics[scale=0.52]{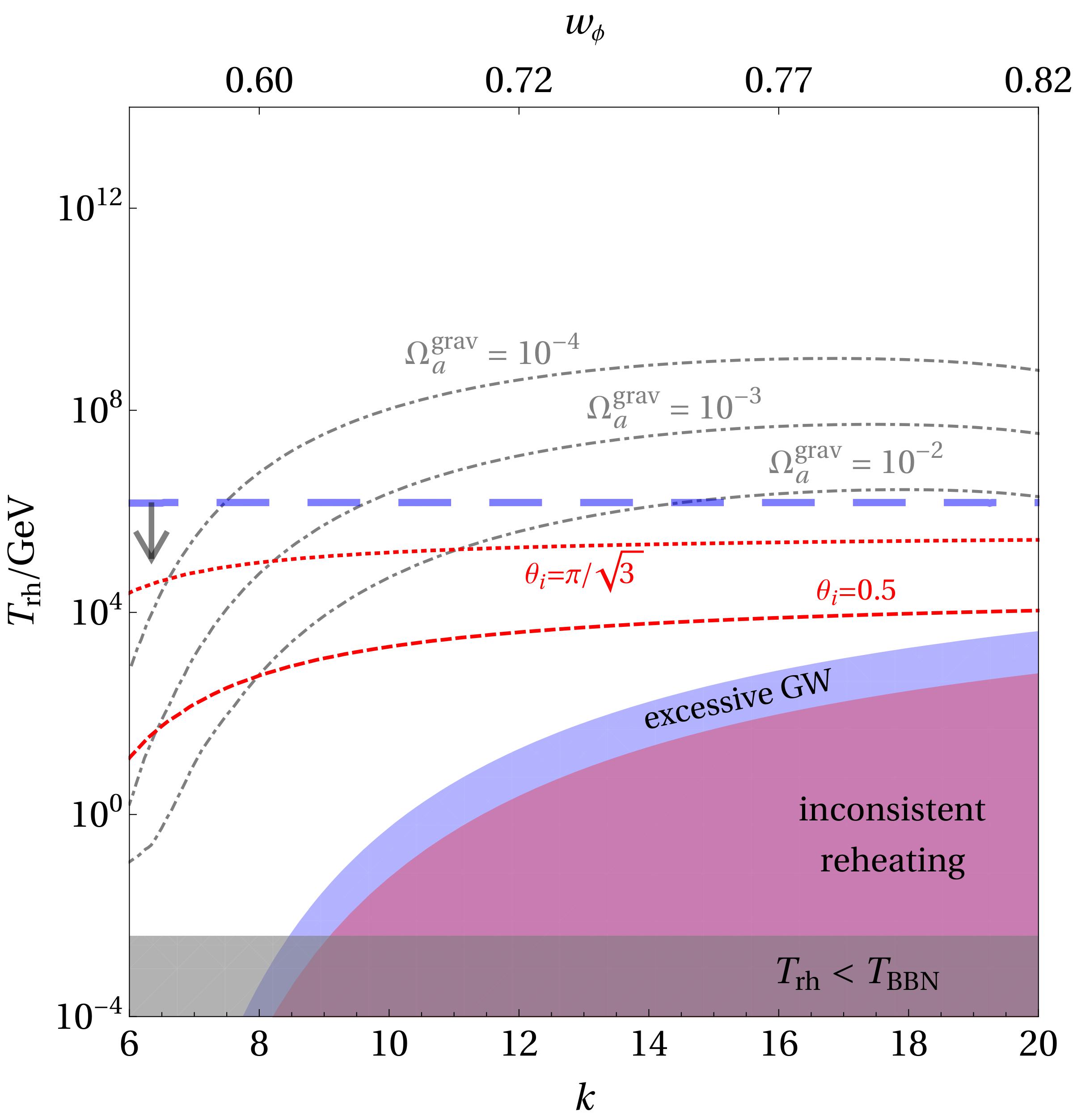}
    \caption{Parameter space in $[\Trh-k]$ plane, where ALP produced via standard misalignment can co-exist with (semi-) relativistic ALPs produced from inflaton scattering during reheating. The gray dashed contours (labeled as $\Omega_a^{\rm grav}$) correspond to underabundance for ALPs produced purely via gravitational scattering. The red dashed contours provide right relic abundance for ALPs produced entirely from misalignment during reheating for $\theta_i=0.5$ (lower dashed contour) and $\theta_i=\pi/\sqrt{3}$ (upper dotted contour). We take $m_a=10^{-2}$ MeV and $f_a=10^{10}$ GeV for all curves. The region of parameter space of our interest $(\Tosc>\Trh)$ is shown by arrowhead.}
    \label{fig:rel-grav}
\end{figure} 
Since the axion number density: $n_a\propto\Trh^2/M_P^3$, in the limit of large $k$ [cf.Eq.~\eqref{eq:nphi}] and $m_a\ll M_P$, for purely gravitational QCD axions $\Omega_a\,h^2\ll 10^{-10}$. This is however not true in case of gravitationally produced ALPs, since their masses can be large enough, independent of the choice of $f_a$. In Fig.~\ref{fig:rel-grav} we show parameter space where the co-existence of misalignment and gravitationally produced population of ALPs can happen for a benchmark value of mass $m_a=0.01$ MeV for scale $f_a=10^{10}$ GeV. For these choice of parameters, ALPs produced via misalignment during reheating can produce the observed relic abundance for $\theta_i=[0.5,\,\pi/\sqrt{3}]$. We therefore show underabundant contours for ALPs produced purely from gravity. We thus see, two populations of QCD axions: one cold, via the standard misalignment mechanism during reheating, and the other originating purely gravitationally from scattering of inflaton can actually co-exist.

\section{Experimental limit on the parameter space}
\label{sec:expt}
In this section, we wish to explore the possibilities of probing the parameter space satisfying the observed relic abundance, for axions produced via standard misalignment during gravitational reheating, at the present and proposed axion search facilities. In order to do that, we examine the interaction of axions with two photons, a highly utilized channel for detecting signatures in both observational studies and experimental investigations. The Lagrangian for such an interaction has the following  form~\cite{Marsh:2015xka,DiLuzio:2020wdo}
\begin{align}
& \mathcal{L}_{a\gamma}\supset -\frac{1}{4}\,g_{a\gamma}\,a\,F^{\mu\nu}\,\tilde{F}_{\mu}=g_{a\gamma}\,{\bf E}\cdot{\bf B}\,,    
\end{align}
where the coupling strength $g_{a\gamma}$ is model dependent and is related to the PQ scale $f_a$ as~\cite{Graham:2015ouw}
\begin{align}
& \left|g_{a\gamma}\right|=\frac{\alpha}{2\,\pi\,f_a}\,\left[\frac{E}{N}-\frac{2}{3}\,\frac{z+4}{z+1}\right]\simeq 10^{-13}\,\text{GeV}^{-1}\,\frac{10^{10}\,\text{GeV}}{f_a}\,,    
\end{align}
where $z=m_u/m_d$ is the ratio of quark masses, and $E$ and $N$ are the electromagnetic and colour anomalies associated with the axion anomaly. For KSVZ models $E/N = 0$~\cite{Kim:1979if,Shifman:1979if}, whereas for DFSZ models $E/N = 8/3$~\cite{Dine:1981rt,Zhitnitsky:1980tq}. For ALPs, on the other hand, one can relate the ALP-photon coupling with the PQ scale $f_a$ via~\cite{Co:2020xlh}
\begin{align}\label{eq:cgamma}
& \left|g_{a\gamma}\right|=\frac{\alpha\,|c_\gamma|}{2\,\pi\,f_a}\,,    
\end{align}
where $|c_\gamma|$ can vary between 1 and 10. We map our viable parameter space on a bi-dimensional plane of $[g_{a\gamma},\,m_a]$, on which we further project limits from different experimental facilities. 
\begin{figure}[htb!]
    \centering
    \includegraphics[scale=0.37]{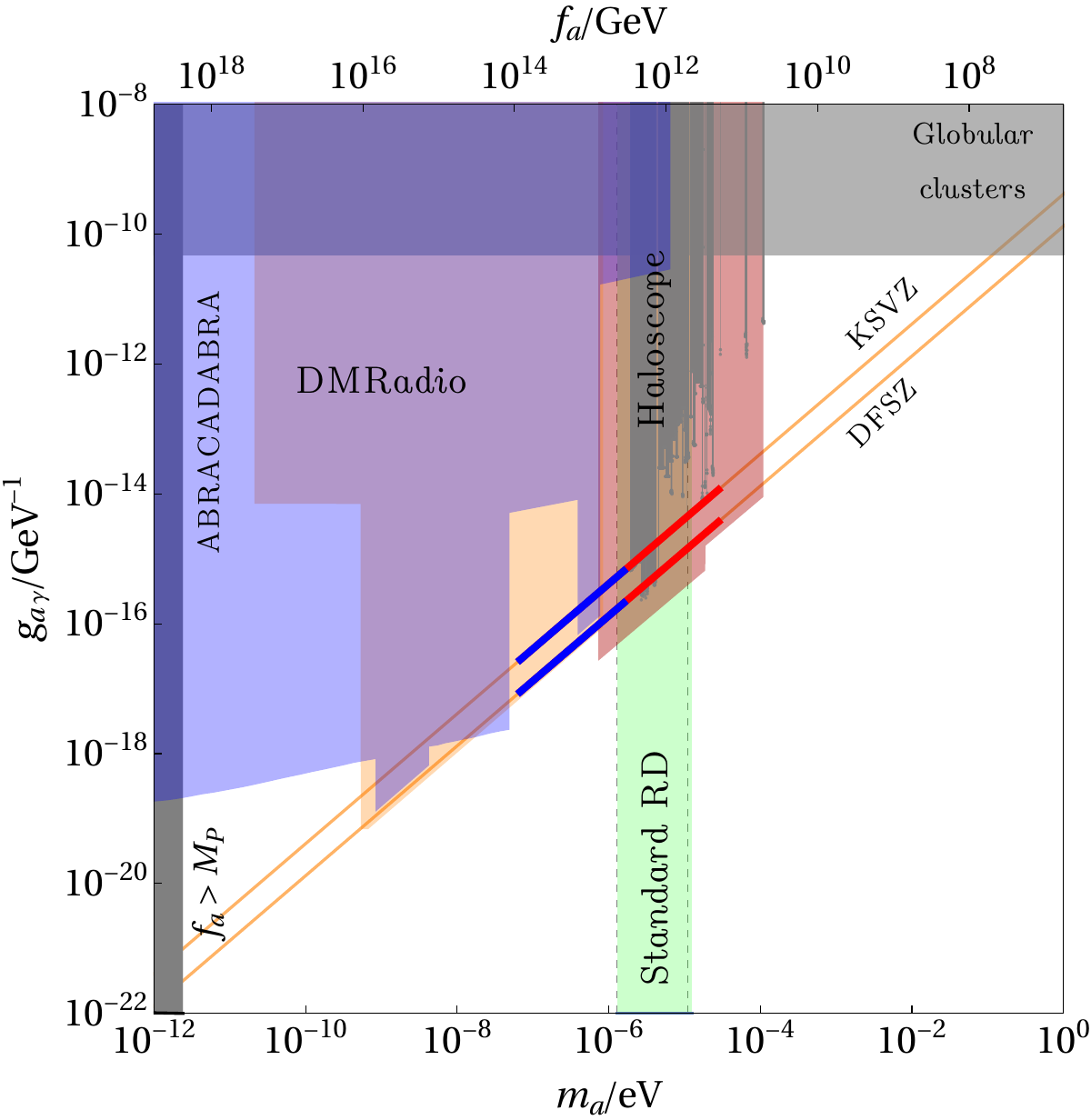}~~\includegraphics[scale=0.37]{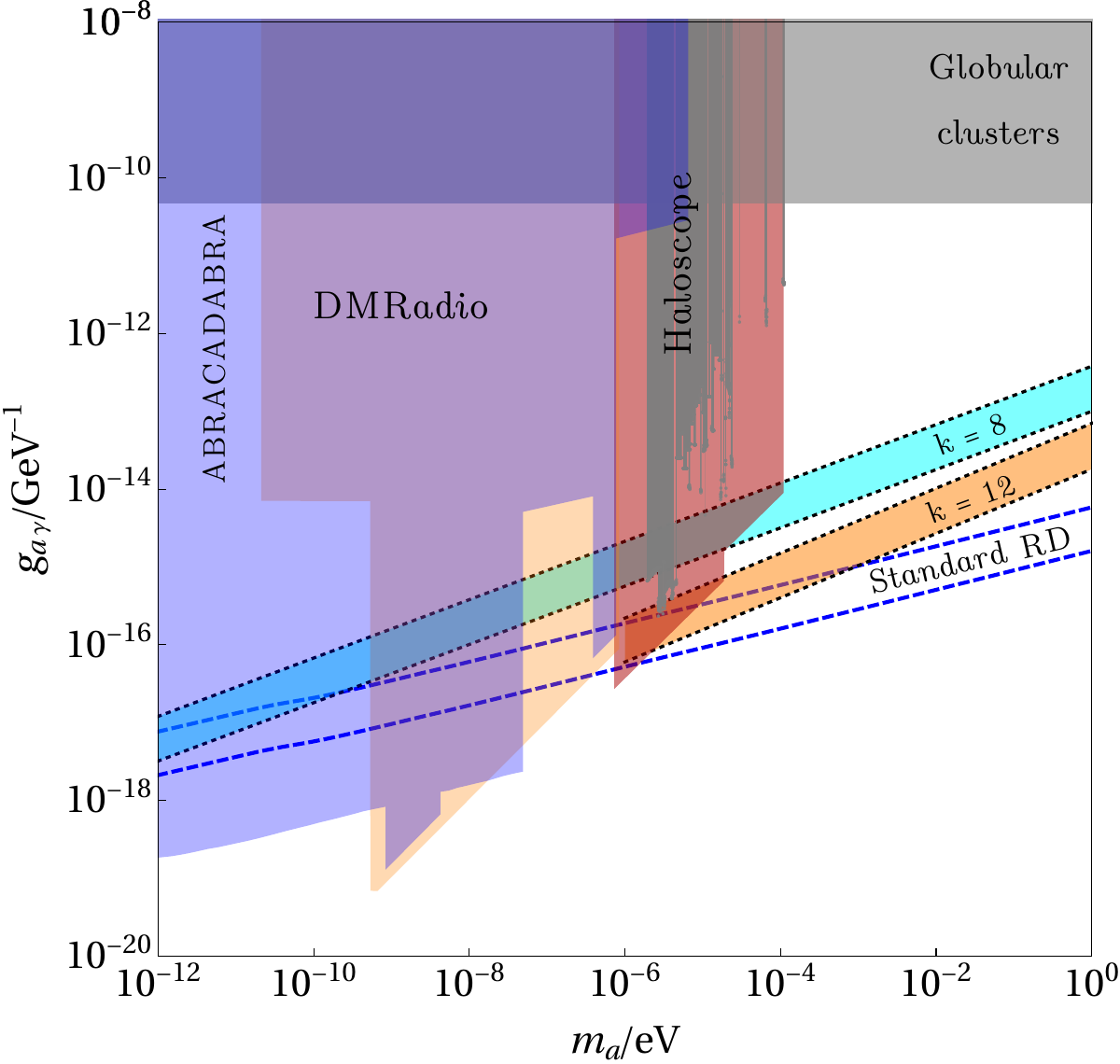}
    \caption{{\it Left:} Parameter space producing right relic abundance, considering gravitational reheating for QCD axion for $E/N=\{0,\,8/3\}$, corresponding to KSVZ and DFSZ model, respectively. The red thick band is disallowed from $\DNeff$ bound from Planck on $\ogw$, while the blue thick band is the viable part of the parameter space. The green shaded band shows axion DM parameter space for standard misalignment during radiation domination. We project limits from a few proposed and existing axion search experiments and astrophysical bounds. {\it Right:} Same as left panel, but for ALPs, where the blue dashed lines correspond to right relic abundance produced from standard misalignment during radiation domination for $|c_\gamma|=1$ and $\theta_i\in[0.5,\,\pi/\sqrt{3}]$. The cyan and orange shaded bands show parameter space where observed DM abundance can be obtained for misalignment during reheating for $k=\{8,\,12\}$, with $\Trh=\{\TBBN,\,10\,\text{GeV}\}$ respectively. In all cases we scan over $\theta_i\in\left[0.5,\,\pi/\sqrt{3}\right]$.}
    \label{fig:expt}
\end{figure} 

In the left panel of Fig.~\ref{fig:expt} we show reach of existing and future experiments in probing the relic density allowed parameter space for QCD axions produced via misalignment during non-minimal gravitational reheating. In obtaining the parameter space, we have scanned over a range of $k\in[6,\,20]$ and $\Trh\in[10^{-4},\,10^{14}]$ GeV along with $\theta_i\in[0.5,\,\pi/\sqrt{3}]$. The vertical green shaded band corresponds to right relic density for QCD axions produced via standard misalignment mechanism during radiation domination. As mentioned before, oscillation during gravitational reheating broadens the window of QCD axion mass (equivalently, the PQ breaking scale) satisfying the observed relic abundance. This is what is reflected here as well. The thick red slanted band is the allowed parameter space that satisfies $\Tosc>\Trh>\TBBN$. However, when the bound $\ogw^{(0)}\,h^2\gtrsim 2\times 10^{-6}$ is imposed on top of that, the parameter space is confined within the blue band. As we see, a part of the viable parameter space is already constrained from existing limits from Haloscope experiments like ADMX~\cite{Asztalos_2010}, CAPP~\cite{Lee:2020cfj}, ORGAN~\cite{McAllister:2017lkb}, as shown by the gray shaded region, while the darker red shaded region is the future projection from Haloscope experiments\footnote{See, for example, Ref.~\cite{Graham:2015ouw} for a review on experimental searches.}. Proposed experiments like DMRadio~\cite{DMRadio:2022pkf} (shown in orange) or future projection from broadband axion-search experiment ABRACADABRA~\cite{Ouellet:2018beu} (shown in blue) are capable of constraining the parameter space further. 

Finally, in the right panel we show relic density allowed parameter space for ALP for two representative values of $k=\{8,\,12\}$, shown via cyan and orange bands, respectively. In this case we consider the entire abundance of ALP is produced through standard misalignment during gravitational reheating. For $k=8$ scenario we choose $\Trh=\TBBN$, whereas for $k=12$, $\Trh\leq 1$ GeV is disallowed from $\DNeff$ bound on GW energy density [cf. Fig.~\ref{fig:Trh}]. We therefore choose $\Trh=10$ GeV corresponding to $k=12$ scenario. In all cases we have considered $|c_\gamma|=1$, and scanned over the initial misalignment angle $\theta_i\in[0.5,\,\pi/\sqrt{3}]$. The first noticeable feature here is that, similar to the case of QCD axions, more parameter space becomes available for ALPs when the misalignment occurs during gravitational reheating. However, since $m_a$ and $f_a$ now can vary independently, hence the resulting parameter space is broader, compared to the QCD axion case. The important point here is to note that for smaller $k$, it is possible to produce the observed abundance with lighter ALPs, a feature we already noticed in the top panel of Fig.~\ref{fig:relALP}. As a result, here we see, with $k=12$, it is possible to obtain the right relic density for $m_a\gtrsim 10^{-15}$ eV, while for masses below this, oscillation happens during RD for $k=12$ and overlaps with the blue dashed curves labelled as ``Standard RD". Another point to note here is that, a lower $\Trh$ requires larger $g_{a\gamma}$ to produce the right abundance for a given mass. This is expected because from Eq.~\eqref{eq:rho-ALP2} we can already see that the relic density goes as $\Omega_a\,h^2\propto 1/\left(g_{a\gamma}\,\Trh^2\right)$, hence the $k=8$ band (with $\Trh=\TBBN$) lies above $k=12$ band (with $\Trh=10$ GeV). Regarding the experimental reach, a part of the parameter space for $k=8$ is already ruled out from present Haloscope experiment, while rest of the parameter space remains well within the reach of future sensitivities from DMRadio, ABRACADABRA and Haloscope experiments. Larger $k$, on  the other hand, is still outside the reach of futuristic experimental facilities as in that case even smaller $g_{a\gamma}$ is required. The precise message here is crucial: any potential discovery of axions in this mass range through futuristic experiments shall not only imply a signature of new physics beyond the SM, but also hint towards gravitational reheating and therefore inflationary paradigm, that may have complementary validation from the detection of primordial gravitational waves in proposed GW detection facilities. 

\section{Conclusions}
\label{sec:concl}
In the simplest scenario, a lone scalar field drives inflation, and the interaction between that scalar field (namely, the inflaton) with the Standard Model (SM) fields is crucial for successful reheating at the end of inflation. However, even without an explicit inflaton-visible sector coupling, gravity-mediated processes can efficiently heat up the Universe after inflation. This gravitational reheating emerges as a minimal and inevitable mechanism for obtaining our current Universe. In addition to the SM particles forming the radiation bath, fields beyond the SM can also be produced through a similar graviton-exchange process, highlighting the democratic nature of gravitational interaction.

With this underlying motivation, in this work we have discussed a scenario where axions, that arise as an elegant solution to the strong-CP problem and can serve as a viable cold dark matter (DM) candidate, are produced via standard misalignment mechanism {\it during} the epoch of reheating. We consider the production of radiation bath purely gravitationally, i.e., from the scattering of inflaton condensate to Higgs final state, mediated by massless graviton. We find, misalignment during gravitational reheating offers a larger window for axion mass, for a natural choice of initial misalignemnt angle $\theta_i\sim\mathcal{O}(1)$, compared to the scenario where misalignment takes place in a radiation-dominated Universe. As the inflaton $\phi$ oscillates in a general monomial potential $\phi^k$, its equation of state mimics that of a stiffer-than-radiation fluid for $k>4$, a value that is anyway required to reheat the Universe prior to BBN via purely gravitational coupling. Such a stiff background equation of state results in a blue-tilted primordial gravitational waves (GW), having inflationary origin, that rules out the {\it minimal} gravitational reheating scenario due to BBN bound on GW energy density, encoded in $\DNeff$. For {\it non-minimal} reheating case, however, this puts a bound on the parameter space satisfying relic abundance, constraining typically smaller $\theta_i$'s (see Fig.~\ref{fig:relic} and \ref{fig:relALP}). Due to irreducible gravitational interaction, apart from standard misalignment, axions are also produced via gravity-mediated scattering of the bath particles and inflaton. Such axions are semi-relativistic in nature, and can be  distinguishable from cold axions produced via misalignment. However, these two populations of axions, namely, (semi-) relativistic and cold, can effectively co-exist (see Fig.~\ref{fig:rel-grav}). Once again, overproduction of primordial GW energy density plays an important role in constraining the resulting parameter space.

We finally discuss the discovery potential of our set-up at present and future axion search frontiers, and find out that Haloscope experiments are quite capable of ruling out and/or constraining the parameter space for both QCD axion and ALPs (see Fig.~\ref{fig:expt}). A large part of the parameter space that is within the reach of (future) Haloscope experiments, can also be probed at several GW detectors, or ruled out from GW overproduction ($\DNeff$ bound). In conclusion, our framework not only provides a complementary avenue for axion searches, but any potential discovery of axions at any of these experiments can also validate a non-standard cosmological epoch during reheating, prior to the onset of BBN.

\section*{Acknowledgements}
The authors would like to thank Nicolás Bernal, Raymond T. Co and Yong Xu for many fruitful discussions, and for providing useful feedback on the draft. AD acknowledges the the National Research Foundation of Korea (NRF) grant funded by the Korean government (2022R1A5A1030700) and the support provided by the Department of Physics, Kyungpook National University.
\bibliographystyle{JHEP}
\bibliography{biblio}

\providecommand{\href}[2]{#2}\begingroup\raggedright\begin{thebibliography}{100}

\bibitem{Weinberg:1977ma}
S.~Weinberg, \emph{{A New Light Boson?}},
  \href{https://doi.org/10.1103/PhysRevLett.40.223}{\emph{Phys. Rev. Lett.}
  {\bfseries 40} (1978) 223}.

\bibitem{Wilczek:1977pj}
F.~Wilczek, \emph{{Problem of Strong $P$ and $T$ Invariance in the Presence of
  Instantons}}, \href{https://doi.org/10.1103/PhysRevLett.40.279}{\emph{Phys.
  Rev. Lett.} {\bfseries 40} (1978) 279}.

\bibitem{Peccei:1977hh}
R.D.~Peccei and H.R.~Quinn, \emph{{CP Conservation in the Presence of
  Instantons}}, \href{https://doi.org/10.1103/PhysRevLett.38.1440}{\emph{Phys.
  Rev. Lett.} {\bfseries 38} (1977) 1440}.

\bibitem{Peccei:1977np}
R.D.~Peccei and H.R.~Quinn, \emph{{Some Aspects of Instantons}},
  \href{https://doi.org/10.1007/BF02730110}{\emph{Nuovo Cim. A} {\bfseries 41}
  (1977) 309}.

\bibitem{Peccei:1977ur}
R.D.~Peccei and H.R.~Quinn, \emph{{Constraints Imposed by CP Conservation in
  the Presence of Instantons}},
  \href{https://doi.org/10.1103/PhysRevD.16.1791}{\emph{Phys. Rev. D}
  {\bfseries 16} (1977) 1791}.

\bibitem{Preskill:1982cy}
J.~Preskill, M.B.~Wise and F.~Wilczek, \emph{{Cosmology of the Invisible
  Axion}}, \href{https://doi.org/10.1016/0370-2693(83)90637-8}{\emph{Phys.
  Lett. B} {\bfseries 120} (1983) 127}.

\bibitem{Stecker:1982ws}
F.W.~Stecker and Q.~Shafi, \emph{{The Evolution of Structure in the Universe
  From Axions}}, \href{https://doi.org/10.1103/PhysRevLett.50.928}{\emph{Phys.
  Rev. Lett.} {\bfseries 50} (1983) 928}.

\bibitem{Dine:1982ah}
M.~Dine and W.~Fischler, \emph{{The Not So Harmless Axion}},
  \href{https://doi.org/10.1016/0370-2693(83)90639-1}{\emph{Phys. Lett. B}
  {\bfseries 120} (1983) 137}.

\bibitem{Abbott:1982af}
L.F.~Abbott and P.~Sikivie, \emph{{A Cosmological Bound on the Invisible
  Axion}}, \href{https://doi.org/10.1016/0370-2693(83)90638-X}{\emph{Phys.
  Lett. B} {\bfseries 120} (1983) 133}.

\bibitem{Arias:2012az}
P.~Arias, D.~Cadamuro, M.~Goodsell, J.~Jaeckel, J.~Redondo and A.~Ringwald,
  \emph{{WISPy Cold Dark Matter}},
  \href{https://doi.org/10.1088/1475-7516/2012/06/013}{\emph{JCAP} {\bfseries
  06} (2012) 013} [\href{https://arxiv.org/abs/1201.5902}{{\ttfamily
  1201.5902}}].

\bibitem{Arvanitaki:2009fg}
A.~Arvanitaki, S.~Dimopoulos, S.~Dubovsky, N.~Kaloper and J.~March-Russell,
  \emph{{String Axiverse}},
  \href{https://doi.org/10.1103/PhysRevD.81.123530}{\emph{Phys. Rev. D}
  {\bfseries 81} (2010) 123530}
  [\href{https://arxiv.org/abs/0905.4720}{{\ttfamily 0905.4720}}].

\bibitem{Adams:2022pbo}
C.B.~Adams et~al., \emph{{Axion Dark Matter}},  in \emph{{Snowmass 2021}}, 3,
  2022 [\href{https://arxiv.org/abs/2203.14923}{{\ttfamily 2203.14923}}].

\bibitem{Co:2019jts}
R.T.~Co, L.J.~Hall and K.~Harigaya, \emph{{Axion Kinetic Misalignment
  Mechanism}},
  \href{https://doi.org/10.1103/PhysRevLett.124.251802}{\emph{Phys. Rev. Lett.}
  {\bfseries 124} (2020) 251802}
  [\href{https://arxiv.org/abs/1910.14152}{{\ttfamily 1910.14152}}].

\bibitem{Chang:2019tvx}
C.-F.~Chang and Y.~Cui, \emph{{New Perspectives on Axion Misalignment
  Mechanism}}, \href{https://doi.org/10.1103/PhysRevD.102.015003}{\emph{Phys.
  Rev. D} {\bfseries 102} (2020) 015003}
  [\href{https://arxiv.org/abs/1911.11885}{{\ttfamily 1911.11885}}].

\bibitem{Barman:2021rdr}
B.~Barman, N.~Bernal, N.~Ramberg and L.~Visinelli, \emph{{QCD Axion Kinetic
  Misalignment without Prejudice}},
  \href{https://doi.org/10.3390/universe8120634}{\emph{Universe} {\bfseries 8}
  (2022) 634} [\href{https://arxiv.org/abs/2111.03677}{{\ttfamily
  2111.03677}}].

\bibitem{Arias:2021rer}
P.~Arias, N.~Bernal, D.~Karamitros, C.~Maldonado, L.~Roszkowski and M.~Venegas,
  \emph{{New opportunities for axion dark matter searches in nonstandard
  cosmological models}},
  \href{https://doi.org/10.1088/1475-7516/2021/11/003}{\emph{JCAP} {\bfseries
  11} (2021) 003} [\href{https://arxiv.org/abs/2107.13588}{{\ttfamily
  2107.13588}}].

\bibitem{Bernal:2021bbv}
N.~Bernal, Y.F.~P\'erez-Gonz\'alez, Y.~Xu and {\'O}.~Zapata, \emph{{ALP dark
  matter in a primordial black hole dominated universe}},
  \href{https://doi.org/10.1103/PhysRevD.104.123536}{\emph{Phys. Rev. D}
  {\bfseries 104} (2021) 123536}
  [\href{https://arxiv.org/abs/2110.04312}{{\ttfamily 2110.04312}}].

\bibitem{Schiavone:2021imu}
F.~Schiavone, D.~Montanino, A.~Mirizzi and F.~Capozzi, \emph{{Axion-like
  particles from primordial black holes shining through the Universe}},
  \href{https://arxiv.org/abs/2107.03420}{{\ttfamily 2107.03420}}.

\bibitem{Arias:2022qjt}
P.~Arias, N.~Bernal, J.K.~Osi\'nski and L.~Roszkowski, \emph{{Dark Matter
  Axions in the Early Universe with a Period of Increasing Temperature}},
  \href{https://arxiv.org/abs/2207.07677}{{\ttfamily 2207.07677}}.

\bibitem{Xu:2023lxw}
Y.~Xu, \emph{{Constraining axion and ALP dark matter from misalignment during
  reheating}}, \href{https://doi.org/10.1103/PhysRevD.108.083536}{\emph{Phys.
  Rev. D} {\bfseries 108} (2023) 083536}
  [\href{https://arxiv.org/abs/2308.15322}{{\ttfamily 2308.15322}}].

\bibitem{Freese:2004vs}
K.~Freese and D.~Spolyar, \emph{{Chain inflation: 'Bubble bubble toil and
  trouble'}}, \href{https://doi.org/10.1088/1475-7516/2005/07/007}{\emph{JCAP}
  {\bfseries 07} (2005) 007}
  [\href{https://arxiv.org/abs/hep-ph/0412145}{{\ttfamily hep-ph/0412145}}].

\bibitem{Freese:2005kt}
K.~Freese, J.T.~Liu and D.~Spolyar, \emph{{Inflating with the QCD axion}},
  \href{https://doi.org/10.1103/PhysRevD.72.123521}{\emph{Phys. Rev. D}
  {\bfseries 72} (2005) 123521}
  [\href{https://arxiv.org/abs/hep-ph/0502177}{{\ttfamily hep-ph/0502177}}].

\bibitem{Pajer:2013fsa}
E.~Pajer and M.~Peloso, \emph{{A review of Axion Inflation in the era of
  Planck}}, \href{https://doi.org/10.1088/0264-9381/30/21/214002}{\emph{Class.
  Quant. Grav.} {\bfseries 30} (2013) 214002}
  [\href{https://arxiv.org/abs/1305.3557}{{\ttfamily 1305.3557}}].

\bibitem{Choi:1994ax}
S.Y.~Choi, J.S.~Shim and H.S.~Song, \emph{{Factorization and polarization in
  linearized gravity}},
  \href{https://doi.org/10.1103/PhysRevD.51.2751}{\emph{Phys. Rev. D}
  {\bfseries 51} (1995) 2751}
  [\href{https://arxiv.org/abs/hep-th/9411092}{{\ttfamily hep-th/9411092}}].

\bibitem{Holstein:2006bh}
B.R.~Holstein, \emph{{Graviton Physics}},
  \href{https://doi.org/10.1119/1.2338547}{\emph{Am. J. Phys.} {\bfseries 74}
  (2006) 1002} [\href{https://arxiv.org/abs/gr-qc/0607045}{{\ttfamily
  gr-qc/0607045}}].

\bibitem{Haque:2022kez}
M.R.~Haque and D.~Maity, \emph{{Gravitational reheating}},
  \href{https://doi.org/10.1103/PhysRevD.107.043531}{\emph{Phys. Rev. D}
  {\bfseries 107} (2023) 043531}
  [\href{https://arxiv.org/abs/2201.02348}{{\ttfamily 2201.02348}}].

\bibitem{Barman:2022qgt}
B.~Barman, S.~Cl\'ery, R.T.~Co, Y.~Mambrini and K.A.~Olive, \emph{{Gravity as a
  portal to reheating, leptogenesis and dark matter}},
  \href{https://doi.org/10.1007/JHEP12(2022)072}{\emph{JHEP} {\bfseries 12}
  (2022) 072} [\href{https://arxiv.org/abs/2210.05716}{{\ttfamily
  2210.05716}}].

\bibitem{Giovannini:1998bp}
M.~Giovannini, \emph{{Gravitational waves constraints on postinflationary
  phases stiffer than radiation}},
  \href{https://doi.org/10.1103/PhysRevD.58.083504}{\emph{Phys. Rev. D}
  {\bfseries 58} (1998) 083504}
  [\href{https://arxiv.org/abs/hep-ph/9806329}{{\ttfamily hep-ph/9806329}}].

\bibitem{Giovannini:1999bh}
M.~Giovannini, \emph{{Production and detection of relic gravitons in
  quintessential inflationary models}},
  \href{https://doi.org/10.1103/PhysRevD.60.123511}{\emph{Phys. Rev. D}
  {\bfseries 60} (1999) 123511}
  [\href{https://arxiv.org/abs/astro-ph/9903004}{{\ttfamily
  astro-ph/9903004}}].

\bibitem{Riazuelo:2000fc}
A.~Riazuelo and J.-P.~Uzan, \emph{{Quintessence and gravitational waves}},
  \href{https://doi.org/10.1103/PhysRevD.62.083506}{\emph{Phys. Rev. D}
  {\bfseries 62} (2000) 083506}
  [\href{https://arxiv.org/abs/astro-ph/0004156}{{\ttfamily
  astro-ph/0004156}}].

\bibitem{Seto:2003kc}
N.~Seto and J.~Yokoyama, \emph{{Probing the equation of state of the early
  universe with a space laser interferometer}},
  \href{https://doi.org/10.1143/JPSJ.72.3082}{\emph{J. Phys. Soc. Jap.}
  {\bfseries 72} (2003) 3082}
  [\href{https://arxiv.org/abs/gr-qc/0305096}{{\ttfamily gr-qc/0305096}}].

\bibitem{Boyle:2007zx}
L.A.~Boyle and A.~Buonanno, \emph{{Relating gravitational wave constraints from
  primordial nucleosynthesis, pulsar timing, laser interferometers, and the
  CMB: Implications for the early Universe}},
  \href{https://doi.org/10.1103/PhysRevD.78.043531}{\emph{Phys. Rev. D}
  {\bfseries 78} (2008) 043531}
  [\href{https://arxiv.org/abs/0708.2279}{{\ttfamily 0708.2279}}].

\bibitem{Stewart:2007fu}
A.~Stewart and R.~Brandenberger, \emph{{Observational Constraints on Theories
  with a Blue Spectrum of Tensor Modes}},
  \href{https://doi.org/10.1088/1475-7516/2008/08/012}{\emph{JCAP} {\bfseries
  08} (2008) 012} [\href{https://arxiv.org/abs/0711.4602}{{\ttfamily
  0711.4602}}].

\bibitem{Li:2021htg}
B.~Li and P.R.~Shapiro, \emph{{Precision cosmology and the stiff-amplified
  gravitational-wave background from inflation: NANOGrav, Advanced LIGO-Virgo
  and the Hubble tension}},
  \href{https://doi.org/10.1088/1475-7516/2021/10/024}{\emph{JCAP} {\bfseries
  10} (2021) 024} [\href{https://arxiv.org/abs/2107.12229}{{\ttfamily
  2107.12229}}].

\bibitem{Artymowski:2017pua}
M.~Artymowski, O.~Czerwinska, Z.~Lalak and M.~Lewicki, \emph{{Gravitational
  wave signals and cosmological consequences of gravitational reheating}},
  \href{https://doi.org/10.1088/1475-7516/2018/04/046}{\emph{JCAP} {\bfseries
  04} (2018) 046} [\href{https://arxiv.org/abs/1711.08473}{{\ttfamily
  1711.08473}}].

\bibitem{Caprini:2018mtu}
C.~Caprini and D.G.~Figueroa, \emph{{Cosmological Backgrounds of Gravitational
  Waves}}, \href{https://doi.org/10.1088/1361-6382/aac608}{\emph{Class. Quant.
  Grav.} {\bfseries 35} (2018) 163001}
  [\href{https://arxiv.org/abs/1801.04268}{{\ttfamily 1801.04268}}].

\bibitem{Bettoni:2018pbl}
D.~Bettoni, G.~Dom\`enech and J.~Rubio, \emph{{Gravitational waves from global
  cosmic strings in quintessential inflation}},
  \href{https://doi.org/10.1088/1475-7516/2019/02/034}{\emph{JCAP} {\bfseries
  02} (2019) 034} [\href{https://arxiv.org/abs/1810.11117}{{\ttfamily
  1810.11117}}].

\bibitem{Figueroa:2019paj}
D.G.~Figueroa and E.H.~Tanin, \emph{{Ability of LIGO and LISA to probe the
  equation of state of the early Universe}},
  \href{https://doi.org/10.1088/1475-7516/2019/08/011}{\emph{JCAP} {\bfseries
  08} (2019) 011} [\href{https://arxiv.org/abs/1905.11960}{{\ttfamily
  1905.11960}}].

\bibitem{Opferkuch:2019zbd}
T.~Opferkuch, P.~Schwaller and B.A.~Stefanek, \emph{{Ricci Reheating}},
  \href{https://doi.org/10.1088/1475-7516/2019/07/016}{\emph{JCAP} {\bfseries
  07} (2019) 016} [\href{https://arxiv.org/abs/1905.06823}{{\ttfamily
  1905.06823}}].

\bibitem{Bernal:2020ywq}
N.~Bernal, A.~Ghoshal, F.~Hajkarim and G.~Lambiase, \emph{{Primordial
  Gravitational Wave Signals in Modified Cosmologies}},
  \href{https://arxiv.org/abs/2008.04959}{{\ttfamily 2008.04959}}.

\bibitem{Ghoshal:2022ruy}
A.~Ghoshal, L.~Heurtier and A.~Paul, \emph{{Signatures of non-thermal dark
  matter with kination and early matter domination. Gravitational waves versus
  laboratory searches}},
  \href{https://doi.org/10.1007/JHEP12(2022)105}{\emph{JHEP} {\bfseries 12}
  (2022) 105} [\href{https://arxiv.org/abs/2208.01670}{{\ttfamily
  2208.01670}}].

\bibitem{Caldwell:2022qsj}
R.~Caldwell et~al., \emph{{Detection of early-universe gravitational-wave
  signatures and fundamental physics}},
  \href{https://doi.org/10.1007/s10714-022-03027-x}{\emph{Gen. Rel. Grav.}
  {\bfseries 54} (2022) 156}
  [\href{https://arxiv.org/abs/2203.07972}{{\ttfamily 2203.07972}}].

\bibitem{Gouttenoire:2021jhk}
Y.~Gouttenoire, G.~Servant and P.~Simakachorn, \emph{{Kination cosmology from
  scalar fields and gravitational-wave signatures}},
  \href{https://arxiv.org/abs/2111.01150}{{\ttfamily 2111.01150}}.

\bibitem{Barman:2023ktz}
B.~Barman, A.~Ghoshal, B.~Grzadkowski and A.~Socha, \emph{{Measuring inflaton
  couplings via primordial gravitational waves}},
  \href{https://doi.org/10.1007/JHEP07(2023)231}{\emph{JHEP} {\bfseries 07}
  (2023) 231} [\href{https://arxiv.org/abs/2305.00027}{{\ttfamily
  2305.00027}}].

\bibitem{Chakraborty:2023ocr}
A.~Chakraborty, M.R.~Haque, D.~Maity and R.~Mondal, \emph{{Inflaton
  phenomenology via reheating in the light of PGWs and latest BICEP/$Keck$
  data}},  \href{https://arxiv.org/abs/2304.13637}{{\ttfamily 2304.13637}}.

\bibitem{Kallosh:2013hoa}
R.~Kallosh and A.~Linde, \emph{{Universality Class in Conformal Inflation}},
  \href{https://doi.org/10.1088/1475-7516/2013/07/002}{\emph{JCAP} {\bfseries
  07} (2013) 002} [\href{https://arxiv.org/abs/1306.5220}{{\ttfamily
  1306.5220}}].

\bibitem{Planck:2018jri}
{\scshape Planck} collaboration, \emph{{Planck 2018 results. X. Constraints on
  inflation}}, \href{https://doi.org/10.1051/0004-6361/201833887}{\emph{Astron.
  Astrophys.} {\bfseries 641} (2020) A10}
  [\href{https://arxiv.org/abs/1807.06211}{{\ttfamily 1807.06211}}].

\bibitem{Giudice:2000ex}
G.F.~Giudice, E.W.~Kolb and A.~Riotto, \emph{{Largest temperature of the
  radiation era and its cosmological implications}},
  \href{https://doi.org/10.1103/PhysRevD.64.023508}{\emph{Phys. Rev. D}
  {\bfseries 64} (2001) 023508}
  [\href{https://arxiv.org/abs/hep-ph/0005123}{{\ttfamily hep-ph/0005123}}].

\bibitem{Garcia:2020wiy}
M.A.G.~Garc\'ia, K.~Kaneta, Y.~Mambrini and K.A.~Olive, \emph{{Inflaton
  Oscillations and Post-Inflationary Reheating}},
  \href{https://doi.org/10.1088/1475-7516/2021/04/012}{\emph{JCAP} {\bfseries
  04} (2021) 012} [\href{https://arxiv.org/abs/2012.10756}{{\ttfamily
  2012.10756}}].

\bibitem{Clery:2021bwz}
S.~Clery, Y.~Mambrini, K.A.~Olive and S.~Verner, \emph{{Gravitational portals
  in the early Universe}},
  \href{https://doi.org/10.1103/PhysRevD.105.075005}{\emph{Phys. Rev. D}
  {\bfseries 105} (2022) 075005}
  [\href{https://arxiv.org/abs/2112.15214}{{\ttfamily 2112.15214}}].

\bibitem{Clery:2022wib}
S.~Clery, Y.~Mambrini, K.A.~Olive, A.~Shkerin and S.~Verner,
  \emph{{Gravitational portals with nonminimal couplings}},
  \href{https://doi.org/10.1103/PhysRevD.105.095042}{\emph{Phys. Rev. D}
  {\bfseries 105} (2022) 095042}
  [\href{https://arxiv.org/abs/2203.02004}{{\ttfamily 2203.02004}}].

\bibitem{Co:2022bgh}
R.T.~Co, Y.~Mambrini and K.A.~Olive, \emph{{Inflationary gravitational
  leptogenesis}},
  \href{https://doi.org/10.1103/PhysRevD.106.075006}{\emph{Phys. Rev. D}
  {\bfseries 106} (2022) 075006}
  [\href{https://arxiv.org/abs/2205.01689}{{\ttfamily 2205.01689}}].

\bibitem{Ichikawa:2008ne}
K.~Ichikawa, T.~Suyama, T.~Takahashi and M.~Yamaguchi, \emph{{Primordial
  Curvature Fluctuation and Its Non-Gaussianity in Models with Modulated
  Reheating}}, \href{https://doi.org/10.1103/PhysRevD.78.063545}{\emph{Phys.
  Rev. D} {\bfseries 78} (2008) 063545}
  [\href{https://arxiv.org/abs/0807.3988}{{\ttfamily 0807.3988}}].

\bibitem{Kainulainen:2016vzv}
K.~Kainulainen, S.~Nurmi, T.~Tenkanen, K.~Tuominen and V.~Vaskonen,
  \emph{{Isocurvature Constraints on Portal Couplings}},
  \href{https://doi.org/10.1088/1475-7516/2016/06/022}{\emph{JCAP} {\bfseries
  06} (2016) 022} [\href{https://arxiv.org/abs/1601.07733}{{\ttfamily
  1601.07733}}].

\bibitem{Ahmed:2022qeh}
A.~Ahmed, B.~Grzadkowski and A.~Socha, \emph{{Higgs Boson-Induced Reheating and
  Dark Matter Production}},
  \href{https://doi.org/10.3390/sym14020306}{\emph{Symmetry} {\bfseries 14}
  (2022) 306}.

\bibitem{Boyle:2005se}
L.A.~Boyle and P.J.~Steinhardt, \emph{{Probing the early universe with
  inflationary gravitational waves}},
  \href{https://doi.org/10.1103/PhysRevD.77.063504}{\emph{Phys. Rev. D}
  {\bfseries 77} (2008) 063504}
  [\href{https://arxiv.org/abs/astro-ph/0512014}{{\ttfamily
  astro-ph/0512014}}].

\bibitem{Watanabe:2006qe}
Y.~Watanabe and E.~Komatsu, \emph{{Improved Calculation of the Primordial
  Gravitational Wave Spectrum in the Standard Model}},
  \href{https://doi.org/10.1103/PhysRevD.73.123515}{\emph{Phys. Rev. D}
  {\bfseries 73} (2006) 123515}
  [\href{https://arxiv.org/abs/astro-ph/0604176}{{\ttfamily
  astro-ph/0604176}}].

\bibitem{Saikawa:2018rcs}
K.~Saikawa and S.~Shirai, \emph{{Primordial gravitational waves, precisely: The
  role of thermodynamics in the Standard Model}},
  \href{https://doi.org/10.1088/1475-7516/2018/05/035}{\emph{JCAP} {\bfseries
  05} (2018) 035} [\href{https://arxiv.org/abs/1803.01038}{{\ttfamily
  1803.01038}}].

\bibitem{Dodelson:1992km}
S.~Dodelson and M.S.~Turner, \emph{{Nonequilibrium neutrino statistical
  mechanics in the expanding universe}},
  \href{https://doi.org/10.1103/PhysRevD.46.3372}{\emph{Phys. Rev. D}
  {\bfseries 46} (1992) 3372}.

\bibitem{Hannestad:1995rs}
S.~Hannestad and J.~Madsen, \emph{{Neutrino decoupling in the early universe}},
  \href{https://doi.org/10.1103/PhysRevD.52.1764}{\emph{Phys. Rev. D}
  {\bfseries 52} (1995) 1764}
  [\href{https://arxiv.org/abs/astro-ph/9506015}{{\ttfamily
  astro-ph/9506015}}].

\bibitem{Dolgov:1997mb}
A.D.~Dolgov, S.H.~Hansen and D.V.~Semikoz, \emph{{Nonequilibrium corrections to
  the spectra of massless neutrinos in the early universe}},
  \href{https://doi.org/10.1016/S0550-3213(97)00479-3}{\emph{Nucl. Phys. B}
  {\bfseries 503} (1997) 426}
  [\href{https://arxiv.org/abs/hep-ph/9703315}{{\ttfamily hep-ph/9703315}}].

\bibitem{Mangano:2005cc}
G.~Mangano, G.~Miele, S.~Pastor, T.~Pinto, O.~Pisanti and P.D.~Serpico,
  \emph{{Relic neutrino decoupling including flavor oscillations}},
  \href{https://doi.org/10.1016/j.nuclphysb.2005.09.041}{\emph{Nucl. Phys. B}
  {\bfseries 729} (2005) 221}
  [\href{https://arxiv.org/abs/hep-ph/0506164}{{\ttfamily hep-ph/0506164}}].

\bibitem{deSalas:2016ztq}
P.F.~de~Salas and S.~Pastor, \emph{{Relic neutrino decoupling with flavour
  oscillations revisited}},
  \href{https://doi.org/10.1088/1475-7516/2016/07/051}{\emph{JCAP} {\bfseries
  07} (2016) 051} [\href{https://arxiv.org/abs/1606.06986}{{\ttfamily
  1606.06986}}].

\bibitem{EscuderoAbenza:2020cmq}
M.~Escudero~Abenza, \emph{{Precision early universe thermodynamics made simple:
  $N_{\rm eff}$ and neutrino decoupling in the Standard Model and beyond}},
  \href{https://doi.org/10.1088/1475-7516/2020/05/048}{\emph{JCAP} {\bfseries
  05} (2020) 048} [\href{https://arxiv.org/abs/2001.04466}{{\ttfamily
  2001.04466}}].

\bibitem{Akita:2020szl}
K.~Akita and M.~Yamaguchi, \emph{{A precision calculation of relic neutrino
  decoupling}},
  \href{https://doi.org/10.1088/1475-7516/2020/08/012}{\emph{JCAP} {\bfseries
  08} (2020) 012} [\href{https://arxiv.org/abs/2005.07047}{{\ttfamily
  2005.07047}}].

\bibitem{Froustey:2020mcq}
J.~Froustey, C.~Pitrou and M.C.~Volpe, \emph{{Neutrino decoupling including
  flavour oscillations and primordial nucleosynthesis}},
  \href{https://doi.org/10.1088/1475-7516/2020/12/015}{\emph{JCAP} {\bfseries
  12} (2020) 015} [\href{https://arxiv.org/abs/2008.01074}{{\ttfamily
  2008.01074}}].

\bibitem{Bennett:2020zkv}
J.J.~Bennett, G.~Buldgen, P.F.~De~Salas, M.~Drewes, S.~Gariazzo, S.~Pastor
  et~al., \emph{{Towards a precision calculation of $N_{\rm eff}$ in the
  Standard Model II: Neutrino decoupling in the presence of flavour
  oscillations and finite-temperature QED}},
  \href{https://doi.org/10.1088/1475-7516/2021/04/073}{\emph{JCAP} {\bfseries
  04} (2021) 073} [\href{https://arxiv.org/abs/2012.02726}{{\ttfamily
  2012.02726}}].

\bibitem{Maggiore:1999vm}
M.~Maggiore, \emph{{Gravitational wave experiments and early universe
  cosmology}}, \href{https://doi.org/10.1016/S0370-1573(99)00102-7}{\emph{Phys.
  Rept.} {\bfseries 331} (2000) 283}
  [\href{https://arxiv.org/abs/gr-qc/9909001}{{\ttfamily gr-qc/9909001}}].

\bibitem{Yeh:2022heq}
T.-H.~Yeh, J.~Shelton, K.A.~Olive and B.D.~Fields, \emph{{Probing physics
  beyond the standard model: limits from BBN and the CMB independently and
  combined}}, \href{https://doi.org/10.1088/1475-7516/2022/10/046}{\emph{JCAP}
  {\bfseries 10} (2022) 046}
  [\href{https://arxiv.org/abs/2207.13133}{{\ttfamily 2207.13133}}].

\bibitem{Abazajian:2019eic}
K.~Abazajian et~al., \emph{{CMB-S4 Science Case, Reference Design, and Project
  Plan}},  \href{https://arxiv.org/abs/1907.04473}{{\ttfamily 1907.04473}}.

\bibitem{CMB-HD:2022bsz}
{\scshape CMB-HD} collaboration, \emph{{Snowmass2021 CMB-HD White Paper}},
  \href{https://arxiv.org/abs/2203.05728}{{\ttfamily 2203.05728}}.

\bibitem{COrE:2011bfs}
{\scshape COrE} collaboration, \emph{{COrE (Cosmic Origins Explorer) A White
  Paper}},  \href{https://arxiv.org/abs/1102.2181}{{\ttfamily 1102.2181}}.

\bibitem{EUCLID:2011zbd}
{\scshape EUCLID} collaboration, \emph{{Euclid Definition Study Report}},
  \href{https://arxiv.org/abs/1110.3193}{{\ttfamily 1110.3193}}.

\bibitem{Lozanov:2016hid}
K.D.~Lozanov and M.A.~Amin, \emph{{Equation of State and Duration to Radiation
  Domination after Inflation}},
  \href{https://doi.org/10.1103/PhysRevLett.119.061301}{\emph{Phys. Rev. Lett.}
  {\bfseries 119} (2017) 061301}
  [\href{https://arxiv.org/abs/1608.01213}{{\ttfamily 1608.01213}}].

\bibitem{Lozanov:2017hjm}
K.D.~Lozanov and M.A.~Amin, \emph{{Self-resonance after inflation: oscillons,
  transients and radiation domination}},
  \href{https://doi.org/10.1103/PhysRevD.97.023533}{\emph{Phys. Rev. D}
  {\bfseries 97} (2018) 023533}
  [\href{https://arxiv.org/abs/1710.06851}{{\ttfamily 1710.06851}}].

\bibitem{Greene:1997fu}
P.B.~Greene, L.~Kofman, A.D.~Linde and A.A.~Starobinsky, \emph{{Structure of
  resonance in preheating after inflation}},
  \href{https://doi.org/10.1103/PhysRevD.56.6175}{\emph{Phys. Rev. D}
  {\bfseries 56} (1997) 6175}
  [\href{https://arxiv.org/abs/hep-ph/9705347}{{\ttfamily hep-ph/9705347}}].

\bibitem{Green:1999yh}
A.M.~Green, \emph{{Supersymmetry and primordial black hole abundance
  constraints}}, \href{https://doi.org/10.1103/PhysRevD.60.063516}{\emph{Phys.
  Rev. D} {\bfseries 60} (1999) 063516}
  [\href{https://arxiv.org/abs/astro-ph/9903484}{{\ttfamily
  astro-ph/9903484}}].

\bibitem{Amin:2011hj}
M.A.~Amin, R.~Easther, H.~Finkel, R.~Flauger and M.P.~Hertzberg,
  \emph{{Oscillons After Inflation}},
  \href{https://doi.org/10.1103/PhysRevLett.108.241302}{\emph{Phys. Rev. Lett.}
  {\bfseries 108} (2012) 241302}
  [\href{https://arxiv.org/abs/1106.3335}{{\ttfamily 1106.3335}}].

\bibitem{Figueroa:2016wxr}
D.G.~Figueroa and F.~Torrenti, \emph{{Parametric resonance in the early
  Universe\textemdash{}a fitting analysis}},
  \href{https://doi.org/10.1088/1475-7516/2017/02/001}{\emph{JCAP} {\bfseries
  02} (2017) 001} [\href{https://arxiv.org/abs/1609.05197}{{\ttfamily
  1609.05197}}].

\bibitem{Garcia:2023eol}
M.A.G.~Garcia and M.~Pierre, \emph{{Reheating after inflaton fragmentation}},
  \href{https://doi.org/10.1088/1475-7516/2023/11/004}{\emph{JCAP} {\bfseries
  11} (2023) 004} [\href{https://arxiv.org/abs/2306.08038}{{\ttfamily
  2306.08038}}].

\bibitem{Garcia:2023dyf}
M.A.G.~Garc\'ia, M.~Gross, Y.~Mambrini, K.A.~Olive, M.~Pierre and J.-H.~Yoon,
  \emph{{Effects of Fragmentation on Post-Inflationary Reheating}},
  \href{https://arxiv.org/abs/2308.16231}{{\ttfamily 2308.16231}}.

\bibitem{GrillidiCortona:2015jxo}
G.~Grilli~di Cortona, E.~Hardy, J.~Pardo~Vega and G.~Villadoro, \emph{{The QCD
  axion, precisely}},
  \href{https://doi.org/10.1007/JHEP01(2016)034}{\emph{JHEP} {\bfseries 01}
  (2016) 034} [\href{https://arxiv.org/abs/1511.02867}{{\ttfamily
  1511.02867}}].

\bibitem{Borsanyi:2016ksw}
S.~Borsanyi et~al., \emph{{Calculation of the axion mass based on
  high-temperature lattice quantum chromodynamics}},
  \href{https://doi.org/10.1038/nature20115}{\emph{Nature} {\bfseries 539}
  (2016) 69} [\href{https://arxiv.org/abs/1606.07494}{{\ttfamily 1606.07494}}].

\bibitem{DiLuzio:2020wdo}
L.~Di~Luzio, M.~Giannotti, E.~Nardi and L.~Visinelli, \emph{{The landscape of
  QCD axion models}},
  \href{https://doi.org/10.1016/j.physrep.2020.06.002}{\emph{Phys. Rept.}
  {\bfseries 870} (2020) 1} [\href{https://arxiv.org/abs/2003.01100}{{\ttfamily
  2003.01100}}].

\bibitem{Kolb:1990vq}
E.W.~Kolb and M.S.~Turner, \emph{{The Early Universe}}, vol.~69 (1990),
  \href{https://doi.org/10.1201/9780429492860}{10.1201/9780429492860}.

\bibitem{Drees:2015exa}
M.~Drees, F.~Hajkarim and E.R.~Schmitz, \emph{{The Effects of QCD Equation of
  State on the Relic Density of WIMP Dark Matter}},
  \href{https://doi.org/10.1088/1475-7516/2015/06/025}{\emph{JCAP} {\bfseries
  06} (2015) 025} [\href{https://arxiv.org/abs/1503.03513}{{\ttfamily
  1503.03513}}].

\bibitem{Hertzberg:2008wr}
M.P.~Hertzberg, M.~Tegmark and F.~Wilczek, \emph{{Axion Cosmology and the
  Energy Scale of Inflation}},
  \href{https://doi.org/10.1103/PhysRevD.78.083507}{\emph{Phys. Rev. D}
  {\bfseries 78} (2008) 083507}
  [\href{https://arxiv.org/abs/0807.1726}{{\ttfamily 0807.1726}}].

\bibitem{Planck:2018vyg}
{\scshape Planck} collaboration, \emph{{Planck 2018 results. VI. Cosmological
  parameters}},
  \href{https://doi.org/10.1051/0004-6361/201833910}{\emph{Astron. Astrophys.}
  {\bfseries 641} (2020) A6}
  [\href{https://arxiv.org/abs/1807.06209}{{\ttfamily 1807.06209}}].

\bibitem{Bernal:2018qlk}
N.~Bernal, M.~Dutra, Y.~Mambrini, K.~Olive, M.~Peloso and M.~Pierre,
  \emph{{Spin-2 Portal Dark Matter}},
  \href{https://doi.org/10.1103/PhysRevD.97.115020}{\emph{Phys. Rev.}
  {\bfseries D97} (2018) 115020}
  [\href{https://arxiv.org/abs/1803.01866}{{\ttfamily 1803.01866}}].

\bibitem{Acero:2008rh}
M.A.~Acero and J.~Lesgourgues, \emph{{Cosmological constraints on a light
  non-thermal sterile neutrino}},
  \href{https://doi.org/10.1103/PhysRevD.79.045026}{\emph{Phys. Rev. D}
  {\bfseries 79} (2009) 045026}
  [\href{https://arxiv.org/abs/0812.2249}{{\ttfamily 0812.2249}}].

\bibitem{Arias:2023wyg}
P.~Arias, N.~Bernal, J.K.~Osi\'nski, L.~Roszkowski and M.~Venegas,
  \emph{{Revisiting signatures of thermal axions in nonstandard cosmologies}},
  \href{https://arxiv.org/abs/2308.01352}{{\ttfamily 2308.01352}}.

\bibitem{Marsh:2015xka}
D.J.E.~Marsh, \emph{{Axion Cosmology}},
  \href{https://doi.org/10.1016/j.physrep.2016.06.005}{\emph{Phys. Rept.}
  {\bfseries 643} (2016) 1} [\href{https://arxiv.org/abs/1510.07633}{{\ttfamily
  1510.07633}}].

\bibitem{Graham:2015ouw}
P.W.~Graham, I.G.~Irastorza, S.K.~Lamoreaux, A.~Lindner and K.A.~van Bibber,
  \emph{{Experimental Searches for the Axion and Axion-Like Particles}},
  \href{https://doi.org/10.1146/annurev-nucl-102014-022120}{\emph{Ann. Rev.
  Nucl. Part. Sci.} {\bfseries 65} (2015) 485}
  [\href{https://arxiv.org/abs/1602.00039}{{\ttfamily 1602.00039}}].

\bibitem{Kim:1979if}
J.E.~Kim, \emph{{Weak Interaction Singlet and Strong CP Invariance}},
  \href{https://doi.org/10.1103/PhysRevLett.43.103}{\emph{Phys. Rev. Lett.}
  {\bfseries 43} (1979) 103}.

\bibitem{Shifman:1979if}
M.A.~Shifman, A.I.~Vainshtein and V.I.~Zakharov, \emph{{Can Confinement Ensure
  Natural CP Invariance of Strong Interactions?}},
  \href{https://doi.org/10.1016/0550-3213(80)90209-6}{\emph{Nucl. Phys. B}
  {\bfseries 166} (1980) 493}.

\bibitem{Dine:1981rt}
M.~Dine, W.~Fischler and M.~Srednicki, \emph{{A Simple Solution to the Strong
  CP Problem with a Harmless Axion}},
  \href{https://doi.org/10.1016/0370-2693(81)90590-6}{\emph{Phys. Lett. B}
  {\bfseries 104} (1981) 199}.

\bibitem{Zhitnitsky:1980tq}
A.R.~Zhitnitsky, \emph{{On Possible Suppression of the Axion Hadron
  Interactions. (In Russian)}}, {\emph{Sov. J. Nucl. Phys.} {\bfseries 31}
  (1980) 260}.

\bibitem{Co:2020xlh}
R.T.~Co, L.J.~Hall and K.~Harigaya, \emph{{Predictions for Axion Couplings from
  ALP Cogenesis}}, \href{https://doi.org/10.1007/JHEP01(2021)172}{\emph{JHEP}
  {\bfseries 01} (2021) 172}
  [\href{https://arxiv.org/abs/2006.04809}{{\ttfamily 2006.04809}}].

\bibitem{Asztalos_2010}
S.J.~Asztalos, G.~Carosi, C.~Hagmann, D.~Kinion, K.~van Bibber, M.~Hotz et~al.,
  \emph{Squid-based microwave cavity search for dark-matter axions},
  \href{https://doi.org/10.1103/physrevlett.104.041301}{\emph{Physical Review
  Letters} {\bfseries 104} (2010) }.

\bibitem{Lee:2020cfj}
S.~Lee, S.~Ahn, J.~Choi, B.R.~Ko and Y.K.~Semertzidis, \emph{{Axion Dark Matter
  Search around 6.7 $\mu$eV}},
  \href{https://doi.org/10.1103/PhysRevLett.124.101802}{\emph{Phys. Rev. Lett.}
  {\bfseries 124} (2020) 101802}
  [\href{https://arxiv.org/abs/2001.05102}{{\ttfamily 2001.05102}}].

\bibitem{McAllister:2017lkb}
B.T.~McAllister, G.~Flower, J.~Kruger, E.N.~Ivanov, M.~Goryachev, J.~Bourhill
  et~al., \emph{{The ORGAN Experiment: An axion haloscope above 15 GHz}},
  \href{https://doi.org/10.1016/j.dark.2017.09.010}{\emph{Phys. Dark Univ.}
  {\bfseries 18} (2017) 67} [\href{https://arxiv.org/abs/1706.00209}{{\ttfamily
  1706.00209}}].

\bibitem{DMRadio:2022pkf}
{\scshape DMRadio} collaboration, \emph{{Projected sensitivity of DMRadio-m3: A
  search for the QCD axion below 1\,\,\ensuremath{\mu}eV}},
  \href{https://doi.org/10.1103/PhysRevD.106.103008}{\emph{Phys. Rev. D}
  {\bfseries 106} (2022) 103008}
  [\href{https://arxiv.org/abs/2204.13781}{{\ttfamily 2204.13781}}].

\bibitem{Ouellet:2018beu}
J.L.~Ouellet et~al., \emph{{First Results from ABRACADABRA-10 cm: A Search for
  Sub-$\mu$eV Axion Dark Matter}},
  \href{https://doi.org/10.1103/PhysRevLett.122.121802}{\emph{Phys. Rev. Lett.}
  {\bfseries 122} (2019) 121802}
  [\href{https://arxiv.org/abs/1810.12257}{{\ttfamily 1810.12257}}].

\end{thebibliography}\endgroup
\end{document}